\documentclass[journal]{IEEEtran}
\usepackage{amsmath}
\usepackage{amssymb}
\DeclareMathOperator{\nn}{{\mathsf{nn}}}
\DeclareMathOperator{\softmax}{{\mathsf{softmax}}}
\DeclareMathOperator{\LSTM}{{\mathsf{LSTM}}}

\newcommand{\etal}{\textit{et al}.}

\usepackage{array}
\usepackage[caption=false]{subfig}
\usepackage{url}
\usepackage{grffile} 
\usepackage{graphicx}

\usepackage{arydshln} 
\usepackage{float}
\usepackage{fancyhdr}

\IEEEoverridecommandlockouts\IEEEpubid{\makebox[\columnwidth]{978-1-5090-1629-7/16/\$31.00~\copyright~2020 IEEE \hfill} \hspace{\columnsep}\makebox[\columnwidth]{ }}

\begin{document}

\title{How to Teach DNNs to Pay Attention to the \\ Visual Modality in Speech Recognition}
%
%
%

\author{George~Sterpu,
        Christian~Saam,
        Naomi~Harte
\thanks{Authors are affiliated with the Department
of Electrical and Electronic Engineering, Trinity College Dublin, Ireland. E-mail: (sterpug@tcd.ie).}%
}

\markboth{}{}

\maketitle

\thispagestyle{fancy}

\begin{abstract}
Audio-Visual Speech Recognition (AVSR) seeks to model, and thereby exploit, the dynamic relationship between a human voice and the corresponding mouth movements.  A recently proposed multimodal fusion strategy, \emph{AV Align}, based on state-of-the-art sequence to sequence neural networks, attempts to model this relationship by explicitly aligning the acoustic and visual representations of speech. This study investigates the inner workings of \emph{AV Align} and visualises the audio-visual alignment patterns. Our experiments are performed on two of the largest publicly available AVSR datasets, TCD-TIMIT and LRS2. We find that \emph{AV Align} learns to align acoustic and visual representations of speech at the frame level on TCD-TIMIT in a generally monotonic pattern. We also determine the cause of initially seeing no improvement over audio-only speech recognition on the more challenging LRS2. We propose a regularisation method which involves predicting lip-related Action Units from visual representations. Our regularisation method leads to better exploitation of the visual modality, with performance improvements between 7\% and 30\% depending on the noise level. Furthermore, we show that the alternative \emph{Watch, Listen, Attend, and Spell} network is affected by the same problem as \emph{AV Align}, and that our proposed approach can effectively help it learn visual representations.
Our findings validate the suitability of the regularisation method to AVSR and encourage researchers to rethink the multimodal convergence problem when having one dominant modality.
\end{abstract}


\IEEEpeerreviewmaketitle

\section{Introduction}
\label{sec:introduction}

\IEEEPARstart{A}{udio}-Visual Speech Recognition (AVSR) is the automatic transcription of spoken utterances using recordings of a person's voice and face  simultaneously. This emulates the natural ability of humans to better understand a spoken message by not only listening to, but also watching, a person talking. Since the benefits of AVSR become more pronounced in adverse environments, it broadens the applicability range of speech recognition technology to busy urban areas with a multitude of noise sources.

The visual modality of speech helps to disambiguate easily confusable sounds, particularly in noisy environments. An example is given in Massaro and Stork~\cite{massaro1998} for the nasal sounds /m/ and /n/, which may sound very similar in noisy conditions, but can be visually distinguished since the lips are closed at onset for /m/, whereas they are open for /n/. The opposite is also true: the sounds /f/ and /v/ look the same on the lips, whereas they can be distinguished acoustically by voicedness (vibration of the vocal folds). Therefore, some features that are hard to distinguish in the acoustic modality can be easily resolved in the visual modality, and vice versa.

The main challenge in AVSR is to sensibly combine the visual and the auditory speech modalities such that the multimodal system surpasses the recognition accuracy of the acoustic system by a sufficiently large margin to justify the increased computational costs. To date, this remains an open research challenge. The upper bound of the visual modality's contribution to ASR performance depends on the linguistic context and noise conditions, and estimating it proves to be more complex than reporting a single number \cite{fernandez2017}. In addition, although human understanding of audio-visual speech perception in the brain has developed considerably in the recent times \cite{oviatt2017, lalor2018, irwin2017}, it is not clear yet how to leverage this knowledge when designing AVSR systems.

A great amount of effort has been spent on visual feature engineering, carefully reviewed in \cite{potamianos2003, FERNANDEZLOPEZ2018, Potamianos:2017}.
Fernandez-Lopez and Sukno~\cite{FERNANDEZLOPEZ2018} review the technologies used in automatic lip-reading for the past ten years, finding that the Discrete Cosine Transform (DCT) coefficients and Active Appearance Model (AAM) parameters are the most popular visual features, while Hidden Markov Models are the most popular classifiers. Lucey~\etal~\cite{Lucey2005} show that DCT features achieve a good speaker separation, but a poor speech separation. Similarly, Cox~\etal~\cite{cox2008} find that AAM features do not extrapolate well to unseen speakers, prompting the authors to call speaker-dependent or multi-speaker recognition evaluations a red herring. Thus the most popular handcrafted visual features are highly speaker dependent and cannot generalise to new conditions.

The paradigm has shifted in recent years, with the increasing availability of data and compute resources. This has allowed researchers to switch focus from finding good visual features to designing neural architectures for learning such representations. Yet, there is no clear consensus on what is an optimal neural network architecture for representing, aligning, and fusing the auditory and visual speech modalities. Two outstanding architectures are the attention-based sequence to sequence network \cite{attention_seq2seq} and the Transformer network \cite{transformers}, with the first using recurrent connections as in Long Short-term Memory (LSTM) cells \cite{lstm}, and the latter being designed with self-attention over feedforward networks. Both architectures share the same encoder-decoder topology, whereby the encoder transforms the input data into abstract representations, and the decoder predicts a grapheme-level output. 


A common problem in AVSR systems based on Deep Neural Networks (DNN) is the faster learning of the sub-task of recognising the acoustic modality, known to be an (almost) unequivocal encoding of the message in clean conditions, which leads to the neglect of the more ambiguous visual modality. 
To address it, Petridis~\etal~\cite{petridis_ctc_2018} and Afouras~\etal~\cite{afouras_pami} train their visual front-ends on a visual sequence classification task requiring knowledge of the word boundaries in a sentence, then store the representations for offline use when decoding text. The word boundaries are inferred using the audio stream, and these timestamps are assumed to have a visual correspondence. Another mitigation strategy is to alternately train the visual and the acoustic front-ends \cite{ICML2011Ngiam_399, chung_cvpr_2017, zhou2018}. When the audio modality is fully masked, the system is forced to lip-read. However, Shillingford~\etal~\cite{shillingford2018} find that decoding graphemes from the visual modality alone proves to be challenging despite training on 3,885 hours of data. Consequently, it is desirable to find an alternative scaffolding approach.


Most DNN-based multimodal systems encode each modality separately, and the representations are fused when decoding \cite{chung_cvpr_2017, petridis_ctc_2018, afouras_pami}. Instead, in the recently proposed \emph{AV Align} \cite{Sterpu_ICMI2018} the acoustic representations of speech are altered by the visual representations during a multimodal encoding process, before decoding starts. In other words, what the system sees influences what it hears. Another distinct feature of \emph{AV Align} is that the alignment is done at every acoustic frame, allowing the encoder representations to be partially reconstructed from the visual signal and limiting the propagation of uncertainties at future timesteps. This allows the learning of the natural asynchrony between sounds and lip movements. Being able to visualise the audio-visual alignments makes the architecture interpretable by design. \emph{AV Align} is a flexible strategy that does not require the features from the two modalities to have identical sampling rates, as in \cite{afouras_pami, petridis_ctc_2018}. These properties, together with the promising results obtained by Sterpu~\etal~\cite{Sterpu_ICMI2018}, suggest that \emph{AV Align} merits a more detailed investigation.



The \emph{AV Align} architecture suffers from the same aforementioned convergence problem of the visual front-end, as hinted in Sterpu~\etal~\cite{Sterpu_ICMI2018} and validated by the experiments in this work.
To address it, we regress Action Units (AUs) \cite{ekman1997face} from the visual representations and introduce an auxiliary loss function on the visual side which is jointly optimised with the character sequence loss of the decoder. Our strategy to regularise the network with a secondary AU loss addresses the convergence problem and enables a performance boost of the audio-visual system on the challenging LRS2 dataset \cite{lrs2}. We demonstrate that it is possible to efficiently train a DNN-based AVSR system with a mere 30 hours of audio-visual data.


This article extends the work of Sterpu~\etal~\cite{Sterpu_ICMI2018} by showing that the AVSR strategy \emph{AV Align} implicitly discovers monotonic alignments between the acoustic and the visual speech representations, which we find to be a necessary condition to improve the speech recognition error rate of the multimodal system. In addition, we investigate the source of divergence on the challenging LRS2 dataset, and propose an architectural improvement to encourage the convergence of the visual front-end in training. Our improved system can now learn speaker and speech independent representations on in-the-wild data. We show that our architectural improvement also applies to the popular \emph{Watch, Listen, Attend, and Spell} (WLAS) network \cite{chung_cvpr_2017}, effectively helping the system learn visual representations and substantially improve the speech recognition performance when compared to the original architecture. This is the only approach in the AVSR field which attempts to model the inherent alignment between the auditory and visual modalities in speech. This approach has further potential in fields outside AVSR that require an alignment between modalities that have time-varying contributions to the overall task.

In Section~\ref{sec:relwork} we analyse the related work in the context of four main challenges specific to AVSR. Section~\ref{sec:av_align} outlines the \emph{AV Align} strategy previously introduced \cite{Sterpu_ICMI2018}, followed by the proposed architectural improvement in Section~\ref{sec:au_loss_method}. In Section~\ref{sec:exp} we present a set of experiments aimed at understanding the roles that the sub-networks play in the process of learning the task and their contributions to the success of the method. We discuss the main implications and conclude with Section~\ref{sec:discuss}, where we describe several challenges to be addressed by future AVSR systems.

\section{Related Work}
\label{sec:relwork}

As pointed out by the study of Harte and Gillen~\cite{tcdtimit}, the audio-visual speech datasets released before 2015 are not suitable for AVSR research for continuous speech, and earlier studies were generally limited in terms of vocabulary size, number of speakers, or utterance complexity. Yet, data complexity is shown to play an important role when training a neural network \cite{arpit2017, zhang2017}. Therefore, showing that a particular neural network architecture performs well on relatively simpler tasks such as closed set utterance classification \cite{petridis_icassp2018} is not a guarantee of success on continuous speech transcription, and we will mainly restrict our review to the large scale approaches. Given the unconventional nature of our contribution to AVSR, it is necessary to introduce some key concepts and discuss how \emph{AV Align} approaches them.

\subsection{Multimodal taxonomy in AVSR}

The recent survey of Baltrusaitis~\etal~\cite{baltrusaitis2019} proposes a new taxonomy for the challenges faced in multimodal machine learning. It goes beyond the traditional pipelines presented in \cite{potamianos2003, Potamianos:2017}, which were mostly limited to feature \emph{extraction} and modality \emph{fusion}, and introduces the \emph{alignment}, \emph{co-learning} and \emph{translation} of modalities, noting that the latter does not represent a challenge in AVSR due to the uniqueness of the label. We consider the related work in AVSR from the perspective of the main challenges identified by \cite{baltrusaitis2019}, as it allows a clearer separation of the proposed techniques. 

\textbf{a) Representation.} Most of the recent work in AVSR uses variations of Convolutional Neural Networks (CNN) to learn visual representations as a function of data, effectively bypassing the necessity for feature design. The review of Purwins~\etal~\cite{purwins2019} shows that the acoustic modality is widely represented as a log mel-scale spectrogram, since learning features directly from time domain signals remains a challenging task. Petridis~\etal~\cite{petridis_icassp2018} find that learning acoustic features directly from the speech waveform outperforms Mel Frequency Cepstral Coefficients (MFCC) in noisy conditions on the simpler task of word classification, yet the authors' subsequent work \cite{petridis_ctc_2018} reverts to MFCC when attempting the more challenging continuous speech recognition, without reporting results with the previously introduced end-to-end architecture, possibly hinting at the difficulties of learning from raw audio.

\textbf{b) Alignment.} Identifying direct relationships between (sub)elements of the visual and auditory modalities is a primary step towards learning enhanced representations. Even when the camera and microphone are time synchronised, there is still a natural asynchrony between sounds and mouth movements \cite{schwartz2014}. Furthermore, Schwartz and Savariaux~\cite{schwartz2014} show that, for chained syllable sequences, the asynchrony fluctuates with the phonetic context, varying between 20 ms audio lead to 70 ms audio lag, and up to 200 ms audio lag (video lead) for more complex speech structures. Karpov~\etal~\cite{karpov2011} also report a variable delay between viseme and phoneme pairs on continuous speech in Russian, noticing a higher visual lead at the start of a sentence, in line with the experiments of Schwartz and Savariaux on isolated syllables. 

The reviews of Potamianos~\etal~\cite{potamianos2003, Potamianos:2017} suggest that modality alignment has generally been overlooked in AVSR. Even the recent work of \cite{petridis_ctc_2018, afouras_pami} rely on the tight synchronicity assumption between the two speech modalities, and merely enforce an identical sampling rate, so that the learnt representations can be conveniently concatenated. An eventual alignment would only happen implicitly, and we anticipate that validating it would take a great effort.
Chung~\etal~\cite{chung_cvpr_2017} propose WLAS, an extension of the sequence to sequence model \cite{attention_seq2seq} for two modalities using two attention mechanisms on the decoder, one for each modality. To the best of our knowledge, this can be thought of as the first attempt at modeling the alignment in neural-based AVSR systems, although this alignment is the indirect result of aligning the output with each input modality. As an alternative to the dual attention decoding design, we \cite{Sterpu_ICMI2018} propose a novel cross-modal alignment architecture, where the acoustic representations are explicitly aligned with the visual ones in an unsupervised way using a cross-modal attention mechanism. Nevertheless, both approaches allow arbitrary sampling rates for each modality. Surprisingly, subsequent work in \cite{afouras_pami}, representing an update of WLAS~\cite{chung_cvpr_2017} according to the authors, proposes a significantly different architecture based on the Transformer network \cite{transformers} which no longer includes an alignment component. This system is trained on the newer LRS2 dataset instead of the original unreleased LRS1 dataset, and there is no published evaluation of the WLAS network on LRS2, making it impossible to draw a direct comparison between the two models proposed by the same group.

Recently, Tao and Busso~\cite{tao_icme2018} introduced an audio-visual feature alignment scheme using an attention mechanism. One LSTM network transforms handcrafted visual features into higher order representations, and a second LSTM processes the acoustic features while also extracting a visual context vector at every frame as a linear combination of all visual representations.
The representations are optimised to minimise the reconstruction error of the acoustic features from the visual features. In contrast, \emph{AV Align} \cite{Sterpu_ICMI2018} can be seen as the end-to-end alternative to \cite{tao_icme2018} with a different objective function: it learns the visual representations directly from the raw pixels, and jointly optimises them with the character-level decoder, minimising the character error rate. \emph{AV Align} is described in this current paper in Section~\ref{sec:av_align}. Regressing audio features from video features as in \cite{tao_icme2018} enables learning from unlabelled data. We argue, however, that rather than learning cross-modal correlations, the network can simply learn to copy audio features to the output, which is encouraged by the reconstruction loss.

\textbf{c) Fusion.} A frequently seen design in neural multimodal fusion involves concatenating time aligned hidden representations from each modality and applying a stack of neural layers to map the representations onto a shared space \cite{petridis_ctc_2018, petridis_icassp2018, Stafylakis2017, afouras_pami}. Instead, the architecture of Chung~\etal~\cite{chung_cvpr_2017} concatenates the visual and auditory context vectors extracted by two independent attention mechanisms. 
The cross-modal attention mechanism of \emph{AV Align} fuses modalities by concatenating the visual context vector with the current state of a multimodal recurrent layer. This is conceptually more closely related to the simpler frame-level concatenation approach, as fusion takes place on the encoder side. The fundamental difference is that each video representation is correlated with the acoustic one through a learnable function, without making use of the timestamps. Zhou~\etal~\cite{zhou2018} propose an update to \cite{chung_cvpr_2017} by incorporating explicit mixing weights for the two context vectors at each timestep. Similarly, using a hybrid DNN-HMM system, \cite{tao_taslp2018} demonstrate the benefit of introducing a gating unit to scale audio-visual features before concatenation. Since the system lacks a modality alignment module, this design may implicitly prefer linguistic units which are already time-synchronised, such as plosives, leading to an under-exploitation of cross-modal correlations, though this hypothesis is not fully explored in that article.

\textbf{d) Co-learning.} When labelled data for a particular task is limited, exploiting unlabelled data in a different modality creates the opportunity to learn more robust representations. Ngiam~\etal~\cite{ICML2011Ngiam_399} explore the cross-modal learning opportunities in greater detail.
They first demonstrate how to learn better visual speech representations given unlabelled audio-visual data for pre-training. In addition, they demonstrate the benefit of learning shared representations which allow cross-modal reconstructions. More recently, transfer learning has gained popularity in AVSR, although it is not as expressive as the two strategies of \cite{ICML2011Ngiam_399}. Transfer learning typically implies pre-training the acoustic and language models on a much larger dataset \cite{chung_cvpr_2017}, or learning visual representations on a word classification task \cite{petridis_ctc_2018, afouras_pami} without fine-tuning for AVSR. None of these pre-training strategies exploit the audio-visual data jointly, and only speed up unimodal representation learning~\cite{He2019} rather than transfer knowledge between modalities. Moreover, building a stronger language model on a large external dataset, as in~\cite{chung_cvpr_2017}, poses the risk of obscuring the true benefit of the visual modality when comparing AVSR methods, and a fair experiment should be designed using an identical amount of text data.
Contrary to leveraging additional external training data at increased cost,
a different school of thought seeks to overcome the fundamental problem of vanishing and exploding gradients with architectural innovations such as gated RNNs~\cite{lstm} and residual connections~\cite{resnet2}. We believe this to be a preferable direction for research in AVSR, and our proposed method is a step in this direction.

\subsection{Lipreading}
In parallel with AVSR, there have been several developments on the continuous visual speech recognition task, also known as sentence lipreading, using either traditional and hybrid systems \cite{Thangthai2017, sterpu2017, Thangthai2018}, or fully neural pipelines \cite{assael2016, shillingford2018, sterpu_icip2018}. Due to the lack of suitable datasets for sentence level AVSR, the terms \emph{lipreading} and \emph{"speech recognition"} have frequently been used in the literature for classifying isolated units from a closed set \cite{petridis_icassp2018, chung2016, wand2016, Stafylakis2017, Stafylakis2018}, and can be misleading to researchers who are not experts in this field. These contributions are similar to the more recent developments for sentence-level lipreading in terms of representations (CNN + RNN), however they typically lack the alignment and decoding components, and are limited to a single prediction per sequence representing the class probabilities of the closed set of words. Adopting the terminology in \cite{purwins2019}, we believe that differentiating between \emph{sequence classification} and \emph{sequence transduction}/\emph{recognition} is an important step towards improving the clarity of the contributions in the AVSR space. On a similar topic, Bear and Taylor~\cite{bear2017} discuss the subtle difference between lipreading and speechreading as a function of the image region being exploited.

\subsection{Video Description}

Whilst the preceding review has considered authors working directly on the AVSR problem, there is much to be learned (potentially) from other domains. On the related task of multimodal video description, Hori~\etal~\cite{hori2017} call the WLAS approach of \cite{chung_cvpr_2017} na\"ive and propose an improved variant. The authors claim that their approach \emph{"is the first to fuse multimodal information using attention between modalities in a neural network"}. This approach is different from ours as it does not use an attention mechanism between the auditory and visual modalities, but is conceptually very similar to the dual attention mechanism of WLAS. The proposed improvement is of the same nature as in \cite{zhou2018} where the two context vectors extracted on the decoder side are scaled by learnable weights before being fused through an explicit summation. In this work we explore several potentially more general fusion functions for the cross-modal attention setup of \emph{AV Align}.

\section{Attention-Based Audio-Visual Alignment}

\subsection{AV Align}
\label{sec:av_align}

\begin{figure*}[t]
    \centering
    \includegraphics[width=1.0\linewidth]{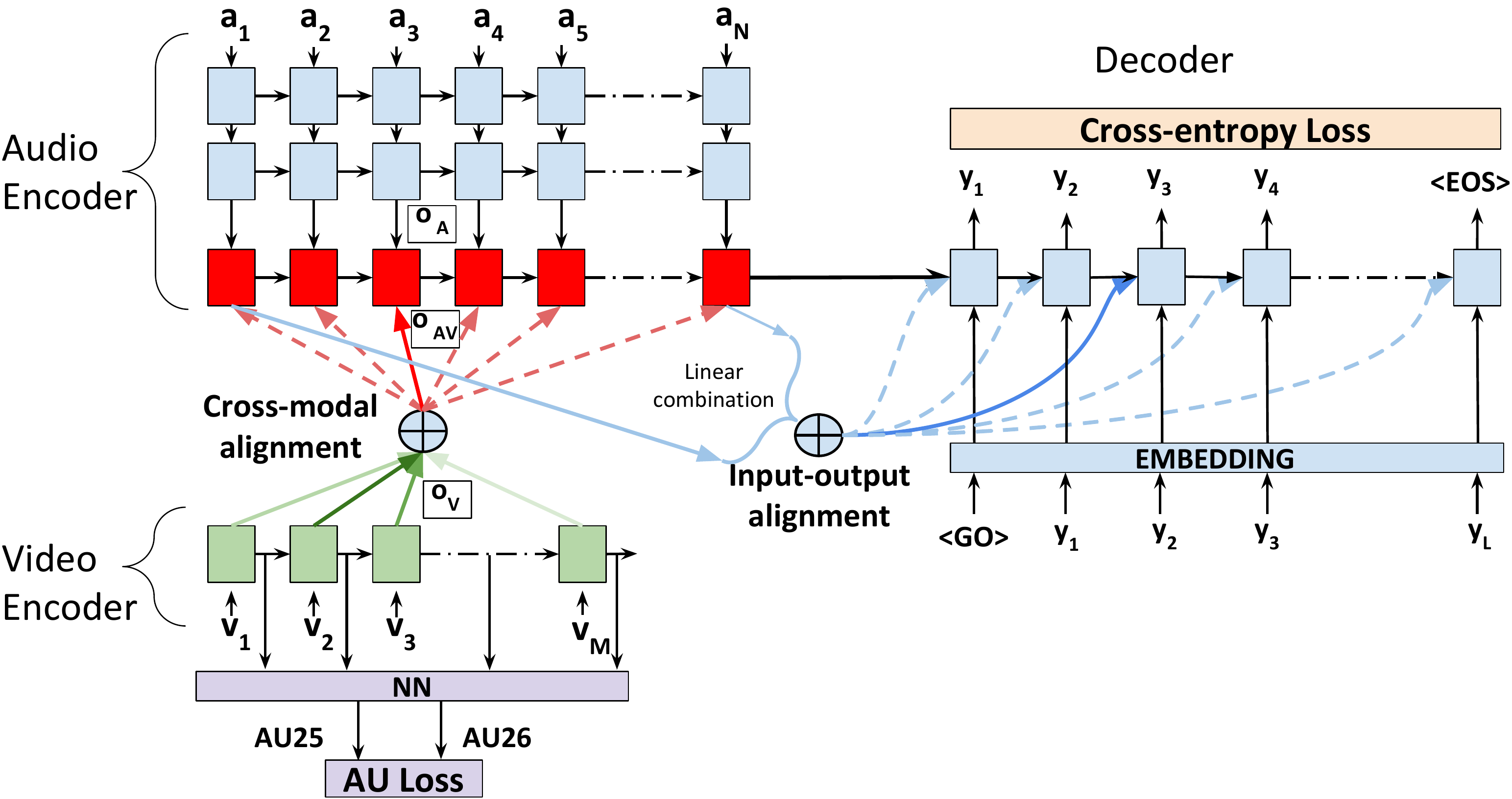}
    \caption{The AV Align strategy. The top layer cells of the Audio Encoder take audio representations from a stack of LSTM layers (${o}_{A}$) as inputs and attend to the top layer outputs of the Video Encoder (${o}_{V}$, only one layer shown), producing the cross-modal alignment. The Decoder receives the fused Audio-Visual representations (${o}_{AV}$), producing an input-output alignment through a second attention mechanism. The architecture in \cite{Sterpu_ICMI2018} is amended with an Action Unit prediction network (\emph{NN + AU Loss}). Dashed lines depict inactive states in a hard selection process, whereas shaded lines stand for a soft selection mechanism.}
    \label{fig:av_align}
\end{figure*}

We review the audio-visual speech fusion strategy \emph{AV Align} originally proposed in \cite{Sterpu_ICMI2018}, and illustrated in Figure~\ref{fig:av_align}. Technically, it can be considered as the original sequence to sequence network with attention \cite{attention_seq2seq} extended with an additional encoder and explicitly modelling the cross-modal correlations.
Given a variable length acoustic sentence $a = \{a_1, a_2, \ldots , a_N\}$ and its corresponding visual track $v = \{v_1, v_2, \ldots , v_M\}$, we transform the raw input signals into higher level latent representations using stacks of LSTM layers (further denoted in Figure~\ref{fig:av_align} by ${o}_{A} = \{o_{A_1}, o_{A_2}, \ldots , o_{A_N}\}$ and ${o}_{V} = \{o_{V_1}, o_{V_2}, \ldots , o_{V_M}\}$):

\begin{align}
    o_{A_i} & = \LSTM_{A} (a_i, o_{A_{i-1}}) \label{eq:oai}\\
    o_{V_j} & = \LSTM_{V} (v_j, o_{V_{j-1}})
\end{align}

Next, one additional LSTM layer is stacked on top of the last acoustic LSTM layer, taking as input ${o}_{A}$. Its hidden state $h_i$ is correlated with all the entries in ${o}_{V}$ at every audio timestep $i$ to compute the visual context vector ${c}_{V_i}$:

\begin{align}
    h_i & = \LSTM_{AV}([o_{A_i}; o_{AV_{i-1}}], h_{i-1}) \label{eq:h_i}\\
    \alpha_{ij} & =  \softmax_i(h_i^{T} \cdot o_{V_j}) \label{eq:alphaij}\\
    \nonumber & \mathsf{where} \ \softmax_i(\mathbf{x}) = \frac{\exp(x_i)}{\sum_j \exp(x_j)} \\
    c_{V_i}  & = \sum_{j=1}^{M} \alpha_{ij}\ \cdot o_{{V}_j}
\end{align}

Finally, the visual context vector $c_{V_i}$ and the current LSTM hidden state $h_i$ are concatenated and fed to a fully-connected neural network having $n$ output units to produce at every timestep the fused representation ${o}_{AV_i}$:

\begin{align}
    o_{AV_i} & = W_{AV} [h_i; c_{V_i}] + b_{AV} \label{eq:oavi}\\
    \nonumber & \mathsf{where} \ W_{AV} \in \mathbb{R}^{\mathrm{n\,x\,2n}}, b_{AV} \in \mathbb{R}^{\mathrm{n}} 
\end{align}

Every input to the attention-enhanced LSTM layer is concatenated with the fused representation from the previous timestep  ${o}_{AV_{i-1}}$, as seen in equation~\eqref{eq:h_i}. Both ${o}_{AV_{0}}$ and $h_0$ are initialised with zeros.

The rest of the network is a character-level LSTM decoder that attends to the enhanced audio-visual representations (${o}_{AV}$) instead of the acoustic only ones (${o}_{A})$, and outputs a variable length character sequence $\hat{y} = \{\hat{y}_1, \hat{y}_2, \ldots, \hat{y}_L\}$. Therefore, \emph{AV Align} adds one cross-modal attention mechanism between the two stream encoders but maintains the traditional attention mechanism between the decoder and encoder.

\begin{align}
    h_{D_k} & = \LSTM_{D}([y_{k-1}; o_{D_{k-1}}], h_{D_{k-1}}) \label{eq:hdk}\\
    \beta_{ki} & = \softmax_k(h_{D_k}^{T} \cdot o_{AV_i}) \label{eq:betakj}\\
    c_{AV_k} & = \sum_{i=1}^{N} \beta_{ki} \cdot o_{AV_i} \\
    o_{D_k} & = W_D [h_{D_k}; c_{AV_k}] + b_D \label{eq:odk}\\
    \nonumber & \mathsf{where} \ W_{D} \in \mathbb{R}^{\mathrm{n\,x\,2n}}, b_{D} \in \mathbb{R}^{\mathrm{n}} \label{eq:last}\\
    p_k & \equiv P(y_{k} | a, v, y_{k-1}) = \softmax(W_{v} o_{D_k} + b_v) \\
    \nonumber & \mathsf{where} \ W_{v} \in \mathbb{R}^{\mathrm{v\,x\,n}}, b_{v} \in \mathbb{R}^{\mathrm{v}} \\
    \hat{y}_k & = \mathrm{arg}\max p_k
\end{align}
and v is the vocabulary size.

Equations~\eqref{eq:h_i}-\eqref{eq:oavi} and \eqref{eq:hdk}-\eqref{eq:odk} represent the default behaviour of the \textit{AttentionWrapper} class in TensorFlow~\cite{abadi2016tensorflow} using the Luong attention mechanism \cite{luong_attention}. The hidden state of the decoder's LSTM layer in equation~\eqref{eq:hdk} is initialised as the final state of the audio-visual LSTM layer: $h_{D_0} = h_N$.

The system is trained using the cross-entropy loss function:

\begin{align}
    CE\ Loss = \frac{1}{L} \sum_k -y_k \log (p_k) \label{eq:celoss}
\end{align}

The motivation behind \emph{AV Align} was to address a possible learning difficulty of the WLAS network, speculating that the dual attention decoder is overburdened with modelling tasks. On top of audio decoding and language modelling, it is also required to learn cross-modal correlations. Instead, \emph{AV Align} moves the cross-modal learning task on the encoder side. Intuitively, the method can be seen as a way of reconstructing and enhancing the frame-level audio representations through the use of a dynamically-computed visual context vector.

\subsection{The Action Unit Loss}
\label{sec:au_loss_method}

As discussed in Section~\ref{sec:introduction}, the original formulation of \emph{AV Align} in \cite{Sterpu_ICMI2018} did not produce satisfactory results on the challenging LRS2 dataset, conflicting with the substantial improvements seen in the controlled conditions of TCD-TIMIT~\cite{tcdtimit}. Initial experiments, reported below in Section~\ref{sec:exp_cross_modal_patterns}, suggest a convergence problem with the visual front-end. Following the reasoning from Section~\ref{sec:introduction}, we want to avoid pre-training strategies and instead rely on the audio-visual data at hand, simplifying the network training methodology.

We suspect that there are two possible causes for the cross-modal attention convergence problem. One is the CNN not learning reliable visual features, as the error signal propagates over a long path susceptible to gradient vanishing. The second one relates to the cross-modal attention mechanism not learning to correlate representations, extract reliable visual context vectors or enhance the acoustic representation. Since the second factor could just be a consequence of the first, we would now like to focus on improving the visual representations.

Our choice is to regress Action Units (AUs) \cite{ekman1997face} from the visual representations and apply an auxiliary loss function penalising the difference between the network's prediction and the targets externally estimated with the OpenFace toolkit \cite{openface2}. We argue that learning to predict \emph{certain} AUs is useful to the visual speech recognition task. The auxiliary loss provides a stronger error signal to the visual encoder than the cross-entropy loss on the decoder side, lessening the effect of gradient vanishing.

    \begin{figure}[H]
        \centering
        \includegraphics[width=0.9\linewidth]{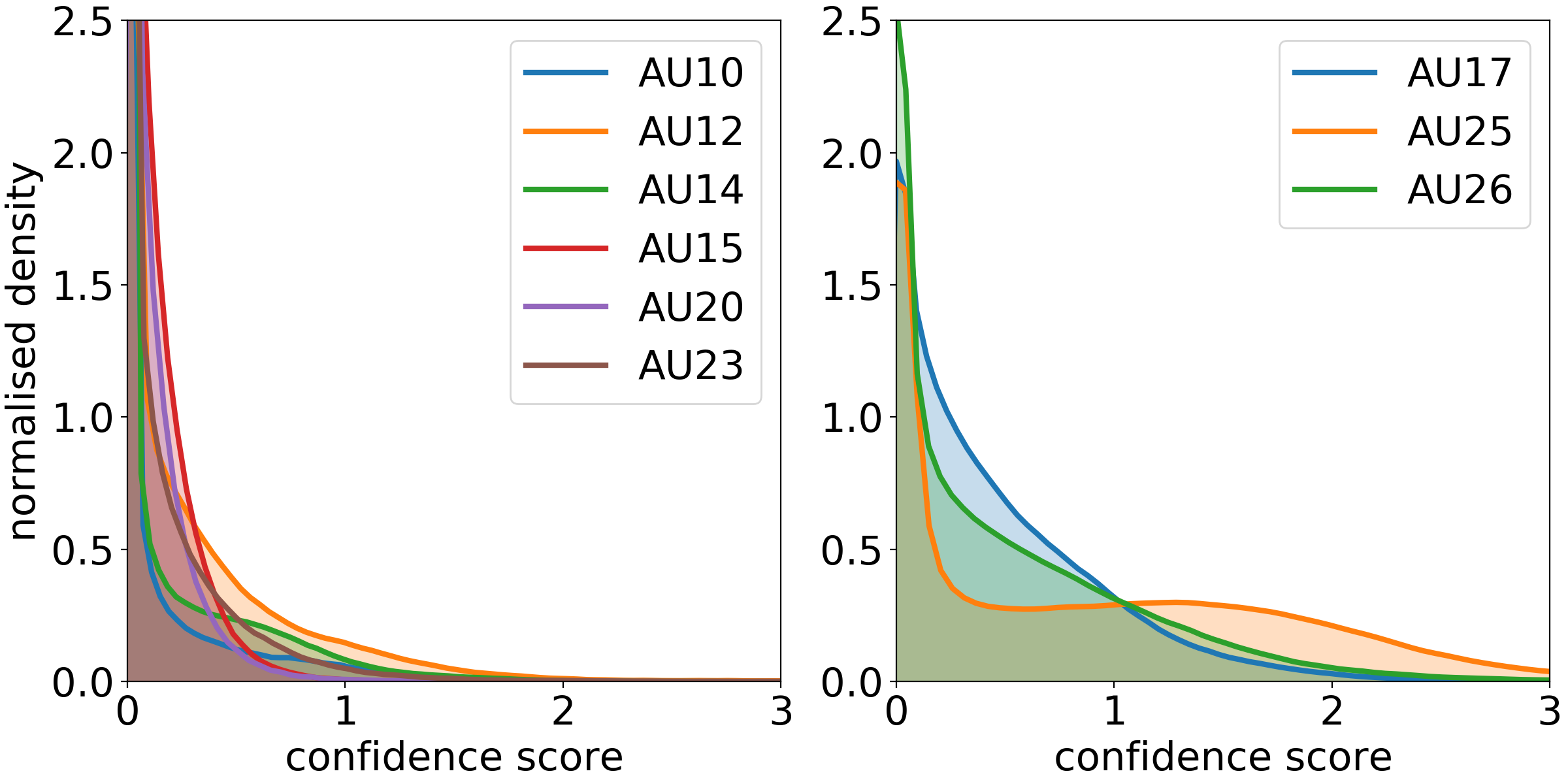}
        \caption{Smoothed histograms of the lip-related Action Units on TCD-TIMIT. Similar histograms are obtained on LRS2.}
        \label{fig:au_histograms}
    \end{figure}

Of the 17 AUs estimated by OpenFace, only 9 refer to the lower face / lip area. A closer inspection of their histograms, displayed in Figure~\ref{fig:au_histograms}, reveals that only three of them (17, 25, 26) occur frequently in speech and could be used for our task. AU17 (\emph{Chin raiser}) appears to be estimated unreliably on our datasets. Consequently we choose only two AUs: \emph{Lips Part} (AU25) and \emph{Jaw Drop} (AU26). These two AUs can be linked to lip opening movements defined in \cite{jeffers1980speechreading} which occur altogether for approximately one third of the time in speech. Although the visibility of the two AUs may be occluded when co-occurring with other action units in speech, estimating the annotations using the video-based OpenFace toolkit ensures that only the visible AUs are taken into account.

We amend the original architecture used in \cite{Sterpu_ICMI2018} with a fully connected neural network (\emph{NN} in Figure~\ref{fig:av_align}, defined in equation~\eqref{eq:au}) with two outputs and sigmoid activation functions, and taking as input the visual LSTM output $o_{V_j}$.
Since AUs are dynamic attributes, we argue that they can be regressed more reliably from $o_{V_j}$, where the temporal context is taken into account, than from the frame level visual features $v_j$.

\begin{align}
    \widehat{AU}_{25, 26} (j) = \mathsf{sigmoid} (W_{AU}\, o_{V_j} + b_{AU}) \label{eq:au}\\
    \nonumber \mathsf{where} \ W_{AU} \in \mathbb{R}^{\mathrm{2\,x\,n}}, b_{AU} \in \mathbb{R}^{\mathrm{2}} \\
    \nonumber \mathsf{and} \ \mathsf{sigmoid}(x) = \frac{1}{1 + \exp(-x)}
\end{align}

To generate the target values, we normalise the intensities estimated with OpenFace, which are real numbers from 0 to 5, by clipping to [0, 3] and dividing by 3 to match the output range of the sigmoid units.
We define the AU Loss function (\emph{AU Loss} in Figure~\ref{fig:av_align}) as the mean squared error between predicted and target AUs, multiplied by a scale factor $\lambda$ of 10.0 found empirically on our evaluation data:

\begin{align}
    AU\ Loss = \frac{\lambda}{M} \sum_{j=1}^{M} (AU_{25, 26} (j) - \widehat{AU}_{25, 26} (j))^2 \label{eq:auloss}
\end{align}

The AU Loss is then added to the decoder's cross entropy loss from equation~\eqref{eq:celoss}.
\section{Experiments and results}
\label{sec:exp}

We begin by presenting the data and the system training procedure, followed by a suite of experiments which offer more insights into the learning mechanisms of \emph{AV Align}.

\subsection{Datasets}

\textbf{TCD-TIMIT} \cite{tcdtimit} consists of high quality audio-visual footage of 62 speakers reading a total of 6,913 examples of both phonetically compact (\emph{sx}) and diverse (\emph{si}) sentences from the prompts of the TIMIT dataset \cite{garofalo1993darpa} in laboratory conditions. It is important to note, in the context of how the results are later discussed, that there is a difference between the coverage of these two types of sentences. Specifically, 450 \emph{sx} sentences are spoken by seven different speakers on average, whereas 1890 \emph{si} sentences are unique to each speaker.
    
\textbf{LRS2} \cite{lrs2} contains 45,839 spoken sentences from BBC television. Unlike TCD-TIMIT, it contains more challenging head poses, uncontrolled illumination conditions, and a much lower image resolution of 160x160 pixels. Both datasets are publicly available for research.

\subsection{Input pre-processing}

Our system takes auditory and visual input concurrently. The \textbf{audio} input is the raw waveform signal of an entire sentence. The \textbf{visual} stream consists of video frame sequences, centred on the speaker's face, which correspond to the audio track. We use the OpenFace toolkit \cite{openface2} to detect and align the faces, then we crop around the lip region.

\textbf{Audio input.} The audio waveforms are re-sampled at 22,050 Hz, and additively mixed with add several types of acoustic noise at different Signal to Noise Ratios (SNR) as explained in Section~\ref{sec:training_proc}. We compute the log magnitude spectrogram of the input, choosing a frame length of 25ms with 10ms stride and 1024 frequency bins for the Short-time Fourier Transform (STFT), and a frequency range from 80Hz to 11,025Hz with 30 bins for the mel scale warp. We stack the features of 8 consecutive STFT frames into a larger window, leading to an audio feature vector $a_i$ of size 240, and we shift this window right by 3 frames, thus attaining an overlap of 5 frames between windows.

\textbf{Visual input.} We down-sample the 3-channel RGB images of the lip regions to 36x36 pixels. A ResNet CNN \cite{resnet2} processes the images to produce a feature vector $v_j$ of \textbf{128 units} per frame. The details of the architecture are presented in Table~\ref{tab:resnet_details}.

\begin{table}[t]
\centering
\caption{CNN Architecture. All convolutions use 3x3 kernels, except the final one. The Residual Block is taken from \cite{resnet2} in its \emph{full preactivation} variant.}
\label{tab:resnet_details}
\begin{tabular}{rcr}
\textbf{layer} & \textbf{operation}                                      & \textbf{output shape} \\ \hline
0              & Rescale [-1 ... +1]                                                & 36x36x\textbf{3}               \\ 
1              & Conv                                                    & 36x36x\textbf{8}               \\ 
2-3            & Res block                                               & 36x36x\textbf{8}               \\ 
4-5            & Res block                                               & 18x18x\textbf{16}              \\ 
6-7            & Res block                                               & 9x9x\textbf{32}                \\ 
8-9            & Res block                                               & 5x5x\textbf{64}                \\ 
10             & Conv 5x5                                                & 1x1x\textbf{128}
\end{tabular}
\end{table}

\subsection{Training procedure}
\label{sec:training_proc}

For our experiments, we train and evaluate audio-only and audio-visual speech recognition models based on the sequence to sequence architecture with attention \cite{attention_seq2seq}. The systems model speech at the character level, with a vocabulary consisting of the 26 letters English alphabet \emph{a--z}, plus blank space and apostrophe. To normalise the text, we convert it to lower case, all numbers are converted to words following the cardinal format, and punctuation is removed. The implementation is done within the publicly available speech recognition toolkit Sigmedia-AVSR \cite{Sterpu_ICMI2018} based on TensorFlow \cite{abadi2016tensorflow}.

Our models and their hyper-parameters are identical to \cite{Sterpu_ICMI2018} except the visual LSTM encoder which was reduced from three to one layer, as an ablation study, not reported here, showed a significant increase of training convergence rate with minimal loss in accuracy. The baseline system consists of a 11-layer ResNet \cite{resnet2} to process the cropped lip images, one or three layers LSTM~\cite{lstm} encoders of 256 units for each modality, and a one-layer LSTM decoder of 256 units. For completeness and reproducibility, we provide the software implementation and all the hyper-parameters at \url{https://github.com/georgesterpu/Sigmedia-AVSR}.

The acoustic modality is corrupted with only \emph{Cafeteria} noise, as this noise type was found the most challenging in \cite{Sterpu_ICMI2018}, and the noise source did not influence the conclusions. We train our systems in four stages, first on clean speech, then with a Signal to Noise Ratio (SNR) of 10db, 0db and finally -5db.
Each time we increment the noise level we also copy the model parameters rather than train from scratch, speeding up the system's convergence.

\subsection{Baseline system performance}

In this section we report the performance of the audio-only and multimodal systems on the two datasets. We repeat each experiment 5 times and report the mean error of the best system, including the 95\% confidence interval displayed as error bars. Additionally, we include the standard deviation of the mean error across the 5 repetitions, displayed with arrows at the bottom of the bar plots.

    \subsubsection{Speaker-Dependent TCD-TIMIT}
    
    We first train Audio-only and Audio-Visual systems on the speaker dependent (SD) partition of TCD-TIMIT, where 70\% of data from 59 speakers is used in training, and the remaining 30\% in evaluation.
    
\begin{figure}[t]
        \centering
        \includegraphics[width=0.9\linewidth]{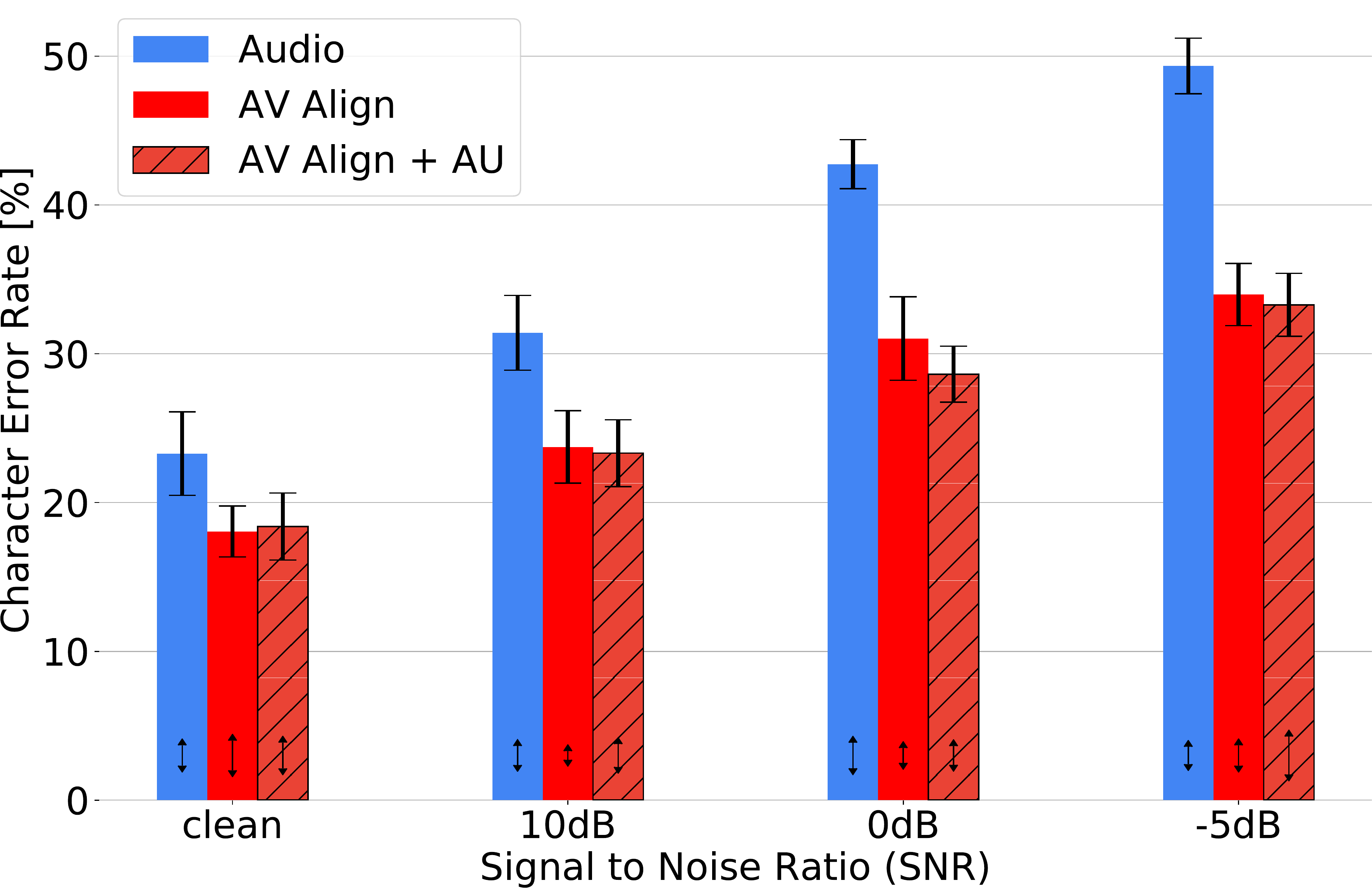}
        \caption{Performance on the Speaker Dependent (SD) partition of TCD-TIMIT.}
        \label{fig:perf_tcd_sd}
    \end{figure}

    Figure~\ref{fig:perf_tcd_sd} shows the Character Error Rate (CER) of our systems for each noise condition. We were able to reproduce the improvements with \emph{AV Align} reported in \cite{Sterpu_ICMI2018}
    across 5 trials, each with a different random initialisation. When we apply the secondary AU loss, the \emph{AV Align + AU} system achieves a similar performance to \emph{AV Align}. This suggests that the AU loss is not detrimental to the performance of \emph{AV Align} when such regularisation is not necessary, as we will show in Section~\ref{sec:exp_cross_modal_patterns}.
    
    A deeper dive into these results reveals that when comparing the audio-visual system with the acoustic-only one, performance gains extend not only to the already seen \emph{sx} sentences, but also to the unique \emph{si} ones. Therefore we can deduce that DNNs can learn sentence independent speech representations. However, it would be much stronger to show that the learnt representations are also speaker independent.
    
    \subsubsection{Speaker-Independent TCD-TIMIT}
    \label{sec:tcd_si_subsec}
    
    We needed a new dataset partitioning scheme to achieve speaker independence. We thus assign each TCD-TIMIT volunteer either to the train or test partition, aiming at the same time to balance attributes such as gender and facial hair. Consequently, speakers \emph{06M, 14M, 17F, 18M, 31F, 41M, 46F, 47M,} and \emph{51F} are assigned to the test set, and the remaining 50 to the train set. Due to the large overlap with the volunteer sentences, the lipspeakers were not used here.
    
\begin{figure}[t]
        \centering
        \includegraphics[width=0.9\linewidth]{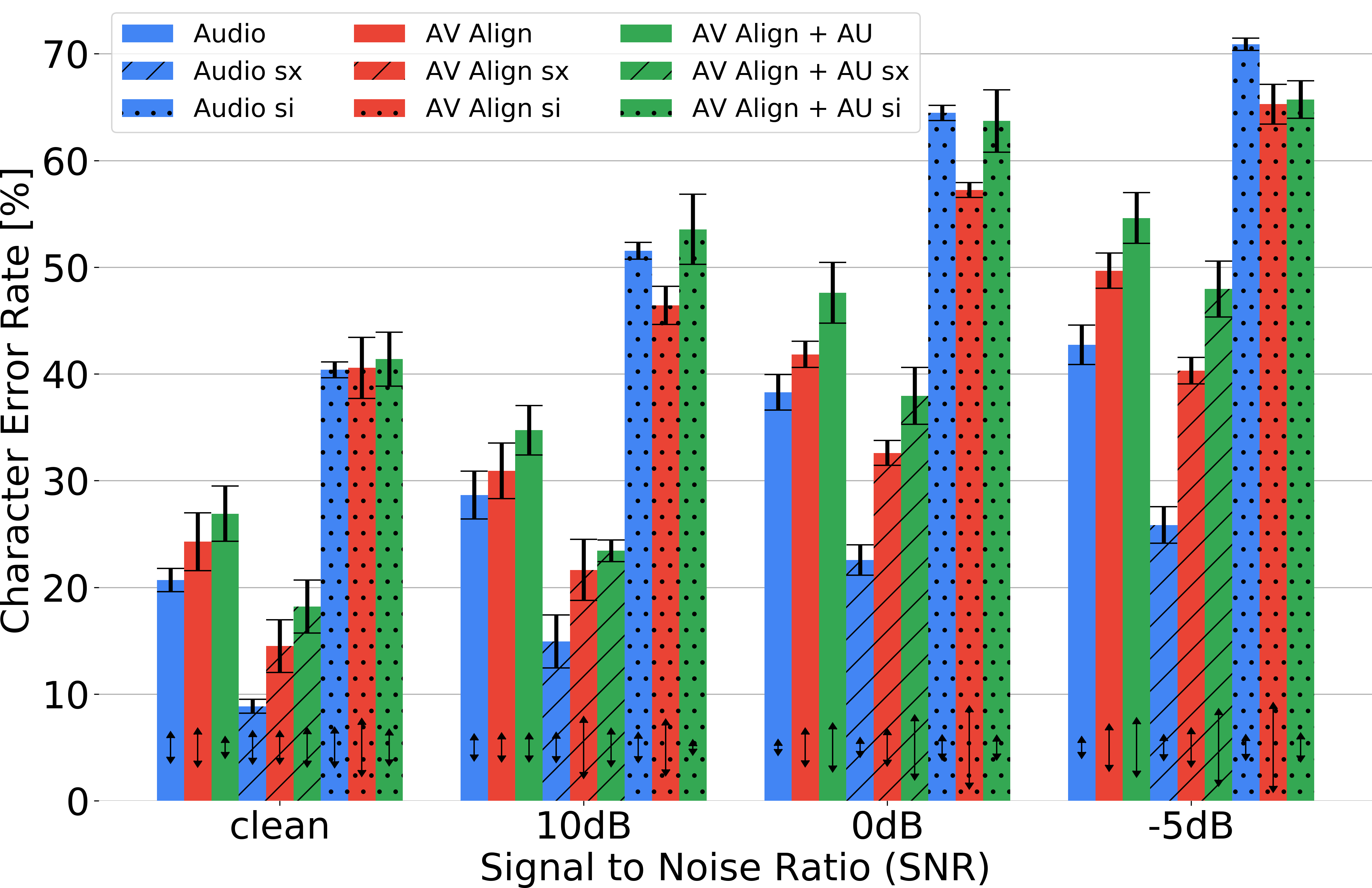}
        \caption{Performance on the Speaker Independent partition of TCD-TIMIT. Note \emph{sx} sentences are repeated across many speakers but \emph{si} sentences are unique to a speaker.}
        \label{fig:tcd_b_perf}
    \end{figure}

    We retrained the audio and audio-visual systems from 5 different random initialisations on this new partition, and display the results in Figure~\ref{fig:tcd_b_perf}, where confidence intervals and standard deviation are displayed as in Figure~\ref{fig:perf_tcd_sd}. Note overall a strong trend whereby performance for the repeated \emph{sx} sentences is markedly better than for the unique \emph{si} sentences. This is apparent in both the audio and audio-visual systems in this speaker-independent scenario.  The global error rate is hence a misleading performance figure. This can likely be attributed to a language model that becomes strongly tuned to the more frequently seen \emph{sx} content due the imbalance between the two sentence types in TCD-TIMIT and the reduced sentence diversity, promoting memorisation~\cite{arpit2017, zhang2017}. Both \emph{AV Align} and \emph{AV Align + AU} frequently converge to poor local optimums where the performance is similar to the \emph{Audio} system. As it will later be shown in Section~\ref{sec:lrs2_alignments}, this corresponds to the case where the audio-visual alignments are not learnt.
    Overall it is difficult to offer definitive conclusions from these experiments. We can see that the variance in performance is reduced by introducing the AU loss, but ultimately it appears we do not have sufficient data for each speaker in TCD-TIMIT to train the speaker-independent system. An ideal dataset would have a much larger number of unique sentences from a larger cohort of speakers, but such a dataset does not exist in the research community. Experiments in the following sections will use LRS2 to allow a fuller exploration of how to optimally exploit the visual modality in speaker-independent AVSR using the \emph{AV Align} strategy.

    \subsubsection{LRS2}
    \label{exp:lrs2_results}
    
\begin{figure}[t]
        \centering
        \includegraphics[width=0.9\linewidth]{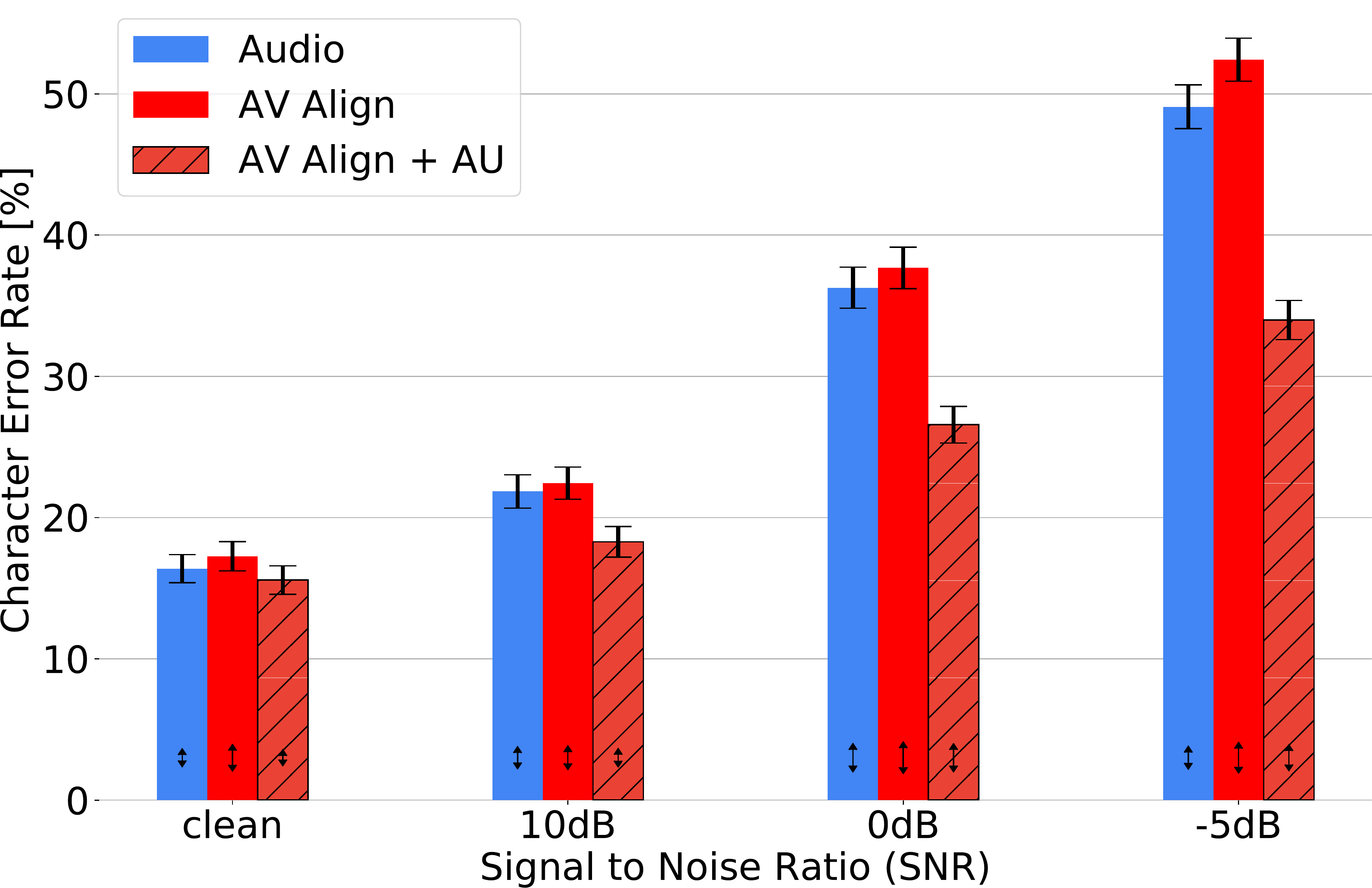}
        \caption{System performance on LRS2}
        \label{fig:lrs2_3}
    \end{figure}

    Since the relatively small size of TCD-TIMIT restricts the learning power of a neural network, as seen in the previous experiment, we now evaluate \emph{AV Align} on LRS2, which is currently the largest publicly available audio-visual dataset. We retrain the audio and audio-visual systems from scratch with the same hyper-parameters, and discard approximately 2.77\% of the LRS2 sentences due to the failures of the face detector. Our results are shown in Figure~\ref{fig:lrs2_3} where confidence intervals and standard deviation are displayed as before.
    
    We notice a relative performance improvement of the \emph{AV Align + AU} system over \emph{Audio} starting at 6.4\% on clean speech, going up to 31\% in worsening audio conditions. These improvements rates had previously only been seen on the speaker dependent partition of TCD-TIMIT, whereas this time we are in the challenging setup of LRS2. Despite the overlap of the 95\% confidence intervals in clean speech, the improvement trend noticed in the remaining noise levels, together with the diminished standard deviation when the AU Loss is used may be indicative of statistical significance even under noise free conditions.
    
    This result brings evidence to support our rationale in Section~\ref{sec:au_loss_method} regarding the difficulties faced by \emph{AV Align} in overcoming the fundamental problem of gradient vanishing. \emph{AV Align} converges to a local minimum where only the audio representations are learnt effectively, and the multitasking design based on the proposed AU Loss was needed so the network could start learning audio and visual representations from the beginning. Nevertheless, these solid improvements on LRS2, as opposed to the inconclusive results in Section~\ref{sec:tcd_si_subsec}, suggest that a strongly imbalanced dataset further contributes to the learning difficulties of the network, and special attention has to be paid to this aspect when collecting new datasets for AVSR.

\subsection{Cross-modal alignment patterns}
\label{sec:exp_cross_modal_patterns}

The \emph{AV Align} architecture allows an explicit soft alignment between the audio and visual representations extracted by the two encoders. A question that arises is: does it \emph{really} learn to align the two modalities of speech, or does it only exploit a spurious correlation in the dataset that would limit the generalisation power? We previously found that the method can decode up to 30\% more accurately than an audio-only system \cite{Sterpu_ICMI2018} on TCD-TIMIT, yet the source of this improvement was unidentified at that time.

     For every sentence in the test set, we generate the alignment matrix between the two encoders, which is the $\alpha_{ij}$ variable in equation~\eqref{eq:alphaij} and has a size of [M x N] corresponding to the number of frames in each modality. Similarly, we also generate the alignment matrix between the decoder state and the fused audio-visual representations, represented by the $\beta_{kj}$ variable in equation~\eqref{eq:betakj} having a size of [N x L] where L is the number of decoded characters. 

    \subsubsection{TCD-TIMIT}

    \begin{figure}[t]
    \centering
    \subfloat[AV alignment on sentence 31F/si459]{
        \includegraphics[width=0.49\linewidth]{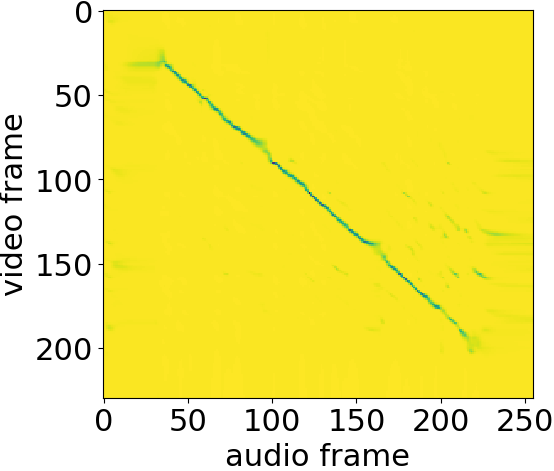}
        \label{fig:alignment_tcd_si459} }
    \subfloat[Mean AV alignment]{
        \includegraphics[width=0.49\linewidth]{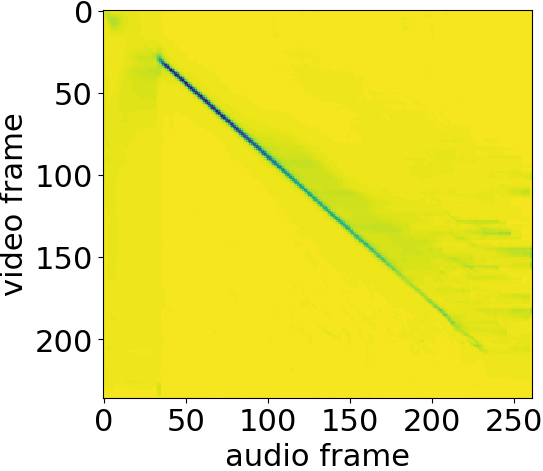}
    \label{fig:alignments_tcd_b_mean} }
    \caption{Cross-modal alignment patterns of the system trained on TCD-TIMIT}
    \end{figure}
    
    We display in Figure~\ref{fig:alignment_tcd_si459} the cross-modal alignment pattern of \emph{AV Align} on a randomly chosen sentence from the speaker dependent test set of TCD-TIMIT. We observe that the alignment pattern looks almost monotonic in a weak sense, i.e. can be well approximated by a monotonic function. The lack of alignment at the start and end of the sentence is attributed to the recording conditions of TCD-TIMIT, where the speakers were instructed to leave a second of silence.
    We also aggregate all the alignments on the test set in Figure~\ref{fig:alignments_tcd_b_mean}, noticing that the monotonicity property is preserved.

    \subsubsection{LRS2}
    \label{sec:lrs2_alignments}
    
\begin{figure}[t]
        \centering
        \subfloat[Audio - Video]{
           \includegraphics[width=0.43\columnwidth]{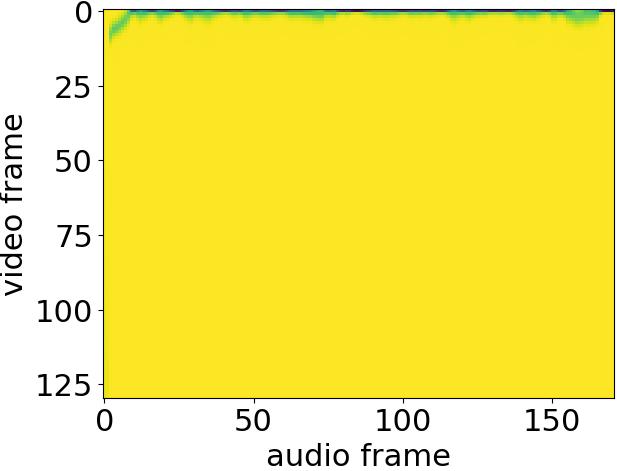}
           \label{fig:alignment_lrs2_av_faulty}
        }
        \subfloat[Decoder - Encoder]{
        \includegraphics[width=0.43\columnwidth]{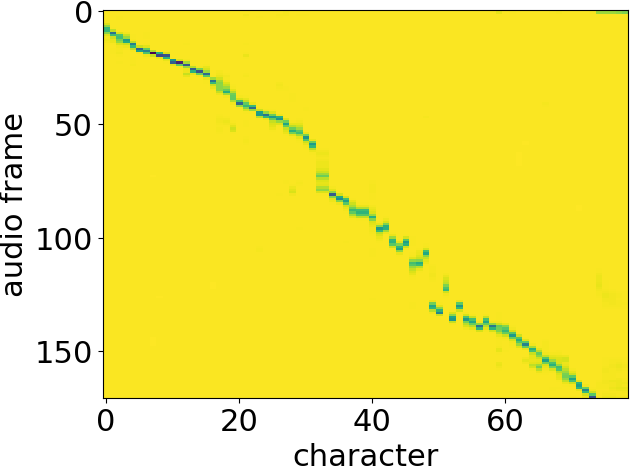}
           \label{fig:alignment_lrs2_at_faulty}
        }
        \caption{Alignment patterns on a single example from LRS2 \hspace{2cm} (6349793037997935601/00008)}
         \label{fig:alignment_lrs2_faulty}
    \end{figure}

     In Figure~\ref{fig:alignment_lrs2_av_faulty} we display the cross-modal alignment patterns of \emph{AV Align} on a randomly chosen example from LRS2, together with the decoder's text alignment in Figure~\ref{fig:alignment_lrs2_at_faulty}.
    
    We observe that each audio frame is predominantly aligned to the first video frame, suggesting a failure of the cross-modal attention mechanism to converge. On the other hand, the second attention mechanism learns non-trivial and plausible alignments between text and inputs. Likely, the fused audio-visual representations are dominated by the audio modality. 
    The performance similarity between \emph{AV Align} and the audio system for all noise levels, illustrated in Figure~\ref{fig:lrs2_3}, brings further evidence to support this claim. We find a similar pattern on the proposed speaker independent partition of TCD-TIMIT.

\begin{figure}[t]
        \centering
        \subfloat[Audio-Video]{
        \includegraphics[width=0.43\columnwidth]{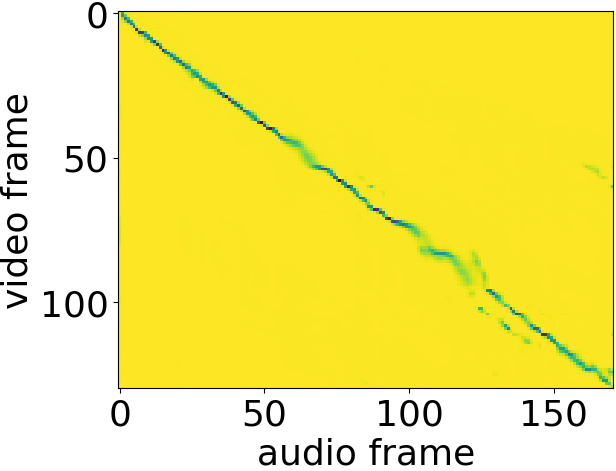}
        \label{fig:exp4_av}
        }
    \subfloat[Decoder - Encoder]{
        \includegraphics[width=0.43\columnwidth]{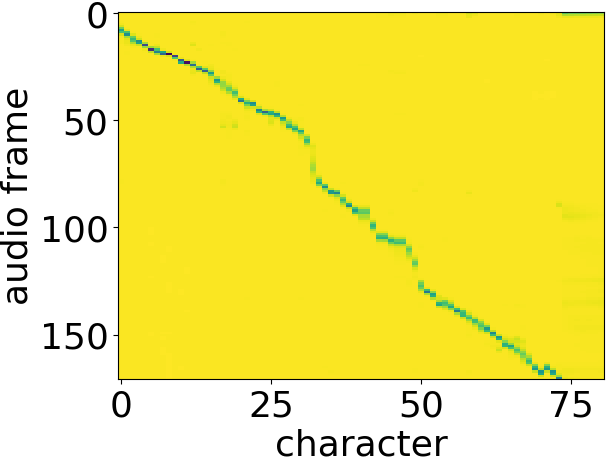}
    \label{fig:exp4_at}
    }
        \caption{Alignment patterns on a single example from LRS2  (6349793037997935601/00008) \textbf{when training with AU Loss}}
        \label{fig:exp4_fixed_alignments}
    \end{figure}

    In Figure~\ref{fig:exp4_fixed_alignments} we display the alignments of \emph{AV Align + AU} on the same sentence as in Figure~\ref{fig:alignment_lrs2_faulty}. This time we see that the system effectively learns proper cross-modal alignments, explaining the performance improvement shown in Figure~\ref{fig:lrs2_3}. Overall, this suggests that monotonic audio-visual alignments are a necessary condition for \emph{AV Align} to capitalise on the visual modality.

\subsection{Additional Control Experiments - Aligning without video}

    To validate that the monotonic alignments represent true correlations between audio and video, we propose three control experiments by corrupting the visual representations $o_{V_j}$ attended to by the audio-visual LSTM layer. These experiments do not require re-training the systems and are only applied for inference.
    
\begin{figure}[t]
        \centering
        \includegraphics[width=0.95\linewidth]{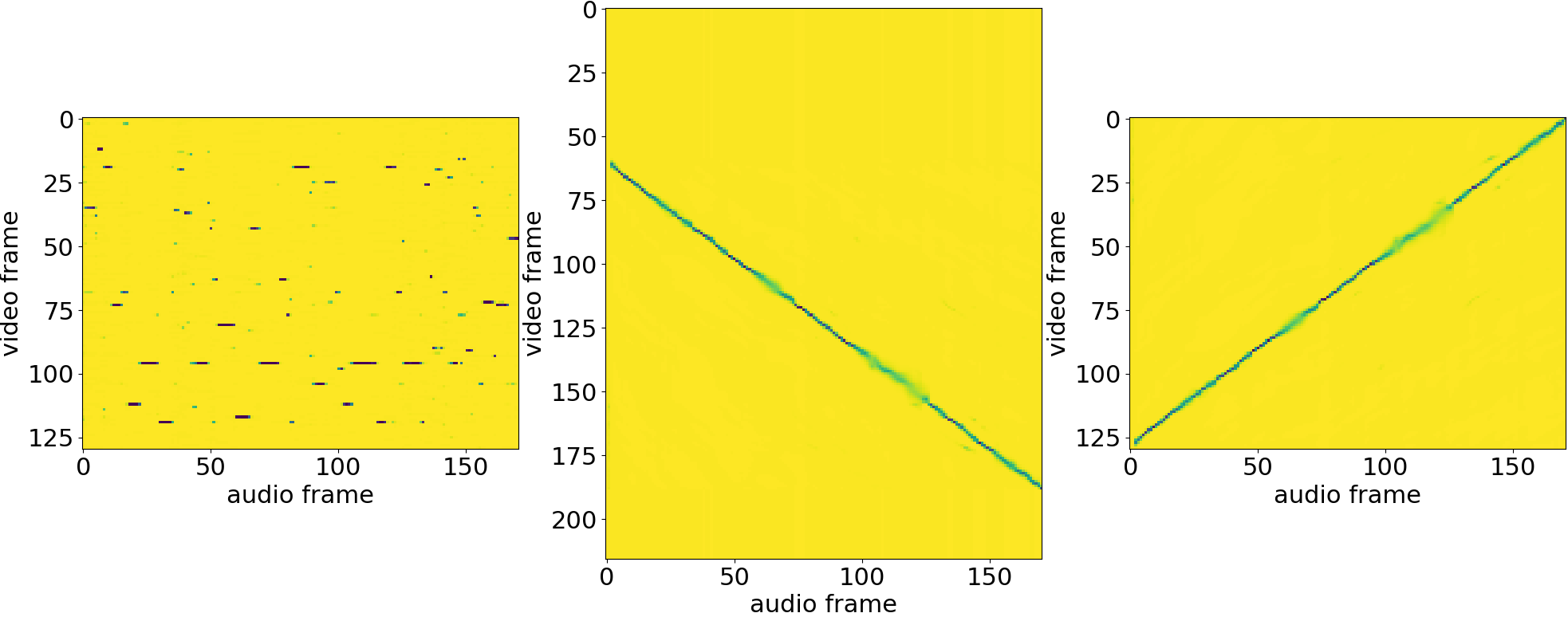}
        
        \hspace{0.8cm}
        \subfloat[]{\label{left} \footnotesize{Random}}
        \hfill
        \subfloat[]{\label{centre} \footnotesize{Blank ends}}
        \hfill
        \subfloat[]{\label{right} \footnotesize{Time reverse}}
        \hspace{0.2cm}
        \hfill

        \caption{The effect of corrupting the video memory with several transformations. Same sentence as in Figure~\ref{fig:exp4_fixed_alignments}.}
        \label{fig:control_experiments}
    \end{figure}

    We first replace the visual representations with random uniform noise. As shown in Figure~\ref{left}, the cross-modal alignment patterns are no longer monotonic as in Figure~\ref{fig:exp4_av}. The error rate surges above 100\%, indicating a limitation of the training strategy to cope with a mismatched data distribution.
    Next, we add segments of blank video frames between one and four seconds long, both in the beginning and at the end of a sentence. We see in Figure~\ref{centre} that the alignment patterns have shifted vertically for a proportional amount of timesteps.
    After reversing the time axis of the video representations, we observe in Figure~\ref{right} that the alignment patterns become horizontally flipped too. 
    The error rate on the test remains identical in this case, whereas it only slightly increases by 0.31\% when appending blank visual frames. These control experiments show that the audio and the visual representations are aligned only by their content, and not because the system implicitly learned to exploit the monotonicity of speech to guess where to look in a sentence.
    
\subsection{Enhancing the representation fusion layer}

    We have shown so far that \emph{AV Align} is able to align audio and video representations, and consequently to fuse them into an informative visual context vector. In the original network \cite{Sterpu_ICMI2018}, the fusion step is implemented as follows: the context vector $c_{V_i}$ is concatenated with the current audio-visual encoder output $h_i$ and processed by a single layer linear neural network. Using the shorthand notation $\nn_{i}(x) = W_i x + b_i$, the fusion function of \cite{Sterpu_ICMI2018}, defined here in equation~\eqref{eq:oavi} was $o_{AV_i} = \nn_{AV}([h_i; c_{V_i}])$ (referred to as \emph{baseline}). We want to explore deeper and nonlinear fusion networks, and we propose the following fusion designs:
    
    {\scriptsize
    \begin{align*}
    M1:\ \ o_{AV_i} & = \tanh(\nn_1(\tanh(\nn_2([h_i; c_{V_i}])))) \\
    M2:\ \ o_{AV_i} & = h_i + \tanh(\nn_1(h_i)) + c_{V_i} + \tanh(\nn_2(c_{V_i})) + \\ & + \tanh(\nn_3([h_i; c_{V_i}])) \\
    M3:\ \ o_{AV_i} & = h_i + \tanh(\nn_1(h_i)) +  \tanh(\nn_2(c_{V_i})) + \\ & + \tanh(\nn_3([h_i; c_{V_i}])) \\
    M4:\ \ o_{AV_i} & = h_i \cdot \mathsf{W_a} + c_{V_i} \cdot \mathsf{W_v}, \mathsf{where} \\
         & \mathsf{W_a} = \mathsf{sigmoid}(\nn_1(h_i)), \\
         & \mathsf{W_v} = \mathsf{sigmoid}(\nn_2(c_{V_i})) \\
    M5:\ \ o_{AV_i} & = h_i \cdot \mathsf{W_a} + \tanh(\nn_1([h_i; c_{V_i}])) \cdot \mathsf{W_{av}}, \\
        & \mathsf{where} \ \mathsf{W_{av}} = \mathsf{sigmoid}(\nn_2([h_i; c_{V_i}]))
    \end{align*}
    }
    
    \vspace{-20pt}
    
    \begin{figure}[t]
        \centering
        \includegraphics[width=0.9\linewidth]{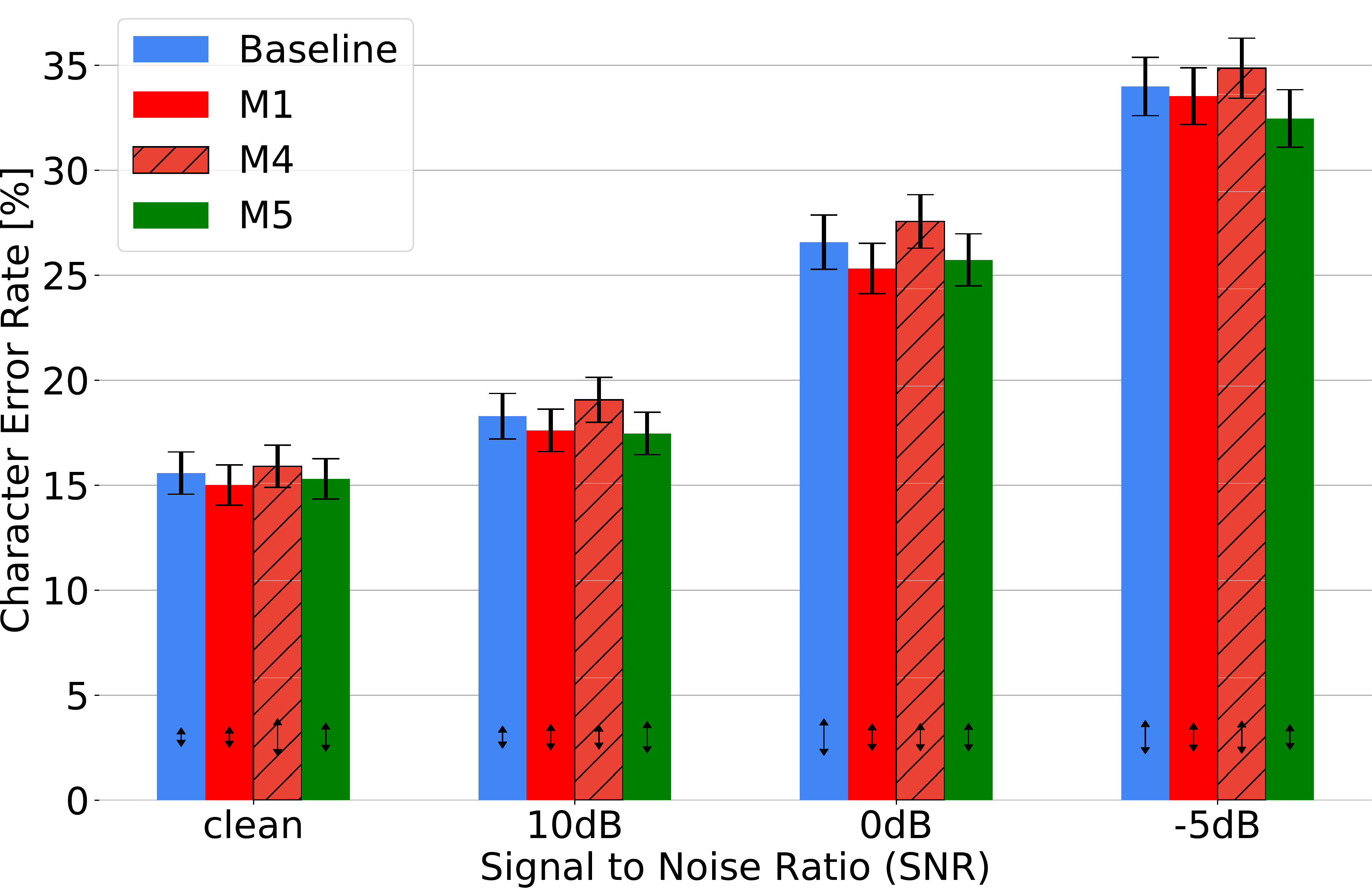}
        \caption{System performance on LRS2 with several variations of the audio-visual fusion layer}
        \label{fig:enhanced_fusion}
    \end{figure}
    
    As can be seen in Figure~\ref{fig:enhanced_fusion}, variants M1 and M5 are relatively up to 5.8\% better than the baseline at certain noise levels. These statistically insignificant improvements suggest either that the original linear fusion is a sufficiently good approximation in AVSR, or that the nonlinearities are learnt by another component of the system, such as the input gate of the LSTM encoder. Interestingly, variant M4 is only 2\% to 4\% worse than the baseline, however it offers a greater interpretability potential since it assigns a confidence score between 0 and 1 to each modality at every timestep. For M1, we also experimented with one and three layer variants using $\mathsf{ReLu}$ and $\mathsf{sigmoid}$ activation functions, all performing slightly below the presented variant. M2 and M3 were similar in performance to M5, and were both omitted from the plot for clarity. Despite not being supported by statistical significance, this experiment illustrates possible extensions of the linear fusion in \emph{AV Align}, which may become useful on larger corpuses.

    \subsection{Generalisation to other AVSR methods}

    In our previous work \cite{Sterpu_ICMI2018}, we could not see a benefit of using the WLAS network \cite{chung_cvpr_2017} over an audio system alone even in the less challenging TCD-TIMIT conditions. Having demonstrated the importance of the AU Loss in Section~\ref{sec:lrs2_alignments} for \emph{AV Align}, an emerging question is: would it also improve WLAS? Our assumption is that the convergence problem is owed to the imbalance in the information carried by the two speech modalities, which is both data and model invariant.
    
\begin{figure}[t]
        \centering
        \includegraphics[width=0.9\linewidth]{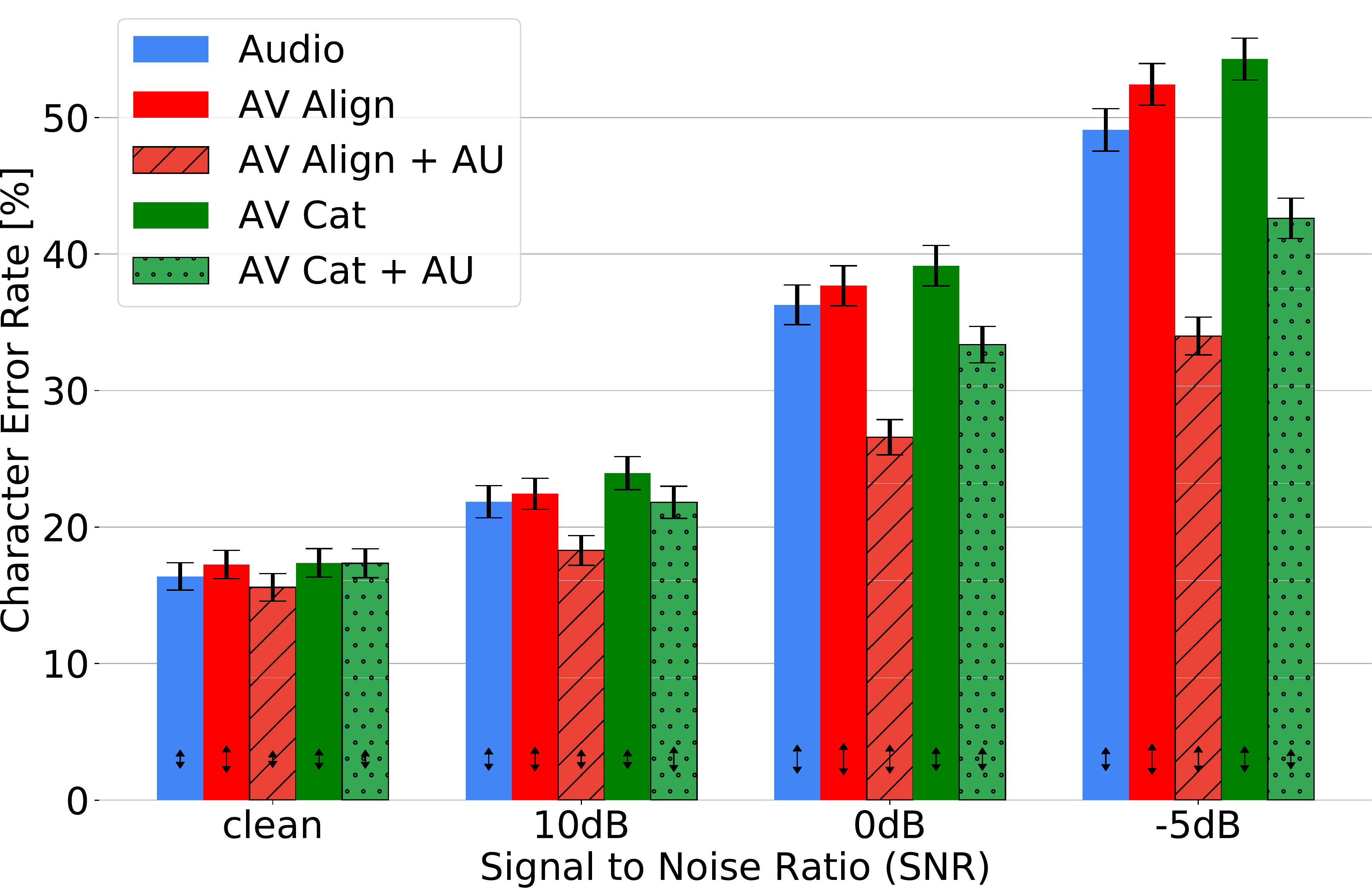}
        \caption{Performance of all five  systems on LRS2}
        \label{fig:lrs2_5}
    \end{figure}

    We implement the WLAS network and follow an identical training and evaluation procedure on LRS2 as with \emph{AV Align}, ensuring a fair comparison. Since we do not pre-train the audio and language models on an auxiliary corpus of 224,528 sentences as in \cite{chung_cvpr_2017}, and we do not make use of the curriculum learning or alternating training,
    essentially training it in the same way as \emph{AV Align}, we denote the network by \emph{AV Cat} instead of \emph{WLAS}. The results are shown in Figure~\ref{fig:lrs2_5}. The original design, \emph{AV Cat}, performs just slightly worse than the audio system, as in \cite{Sterpu_ICMI2018}. The improved model using the AU Loss, \emph{AV Cat + AU}, outperforms the audio model, however its performance is still inferior to \emph{AV Align + AU}.
    
\begin{figure}[t]
        \centering
        \includegraphics[width=0.9\linewidth]{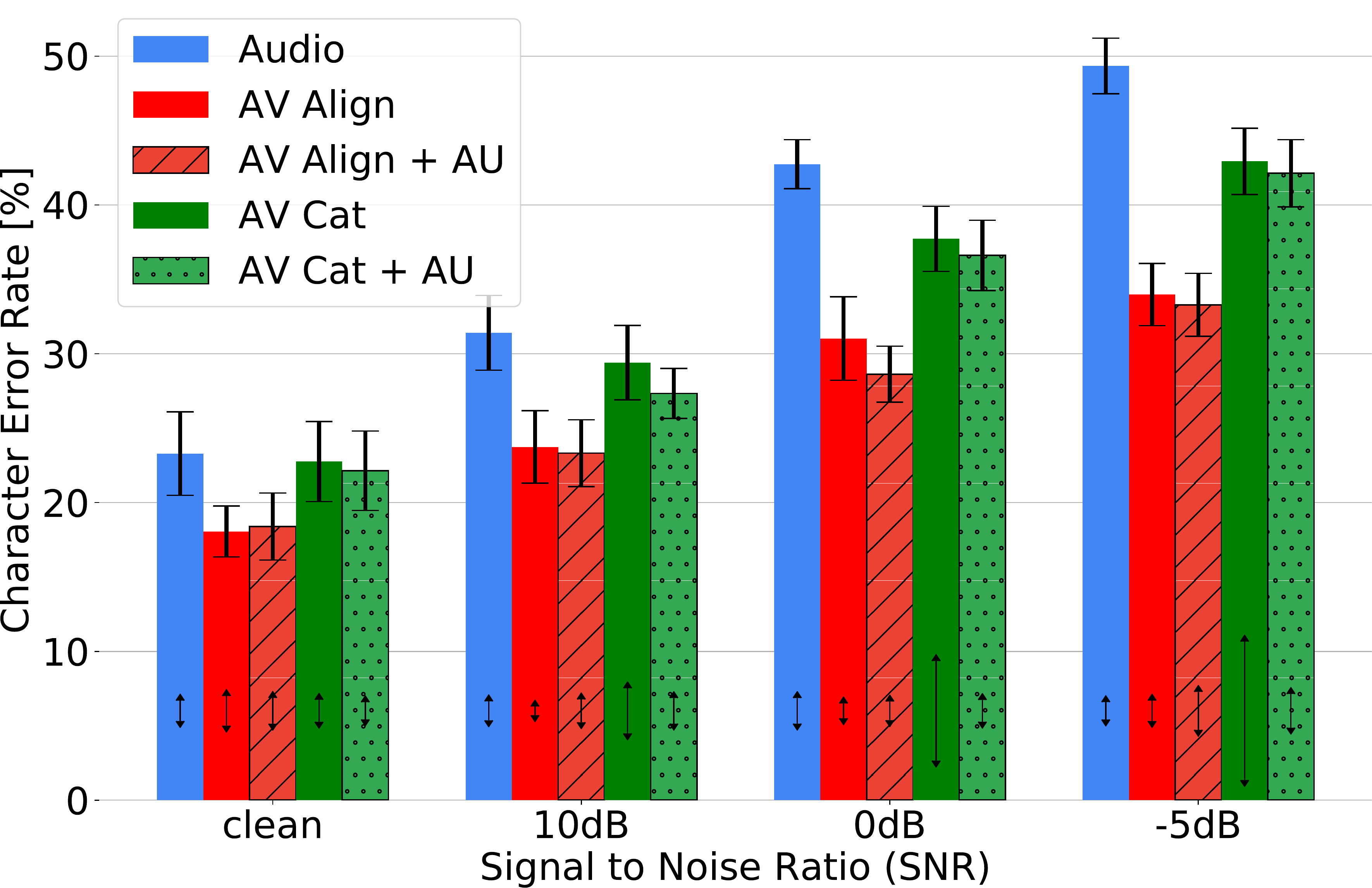}
        \caption{Performance of all five  systems on the SD partition of TCD-TIMIT }
        \label{fig:perf_tcd_sd_5}
    \end{figure}

    Additionally, we train \emph{AV Cat} and \emph{AV Cat + AU} on the SD partition of TCD-TIMIT, and we show the results in Figure~\ref{fig:perf_tcd_sd_5}. The same trend can be seen as in the case of LRS2.
    
\begin{figure}[t]
        \centering
        \includegraphics[width=1.0\linewidth]{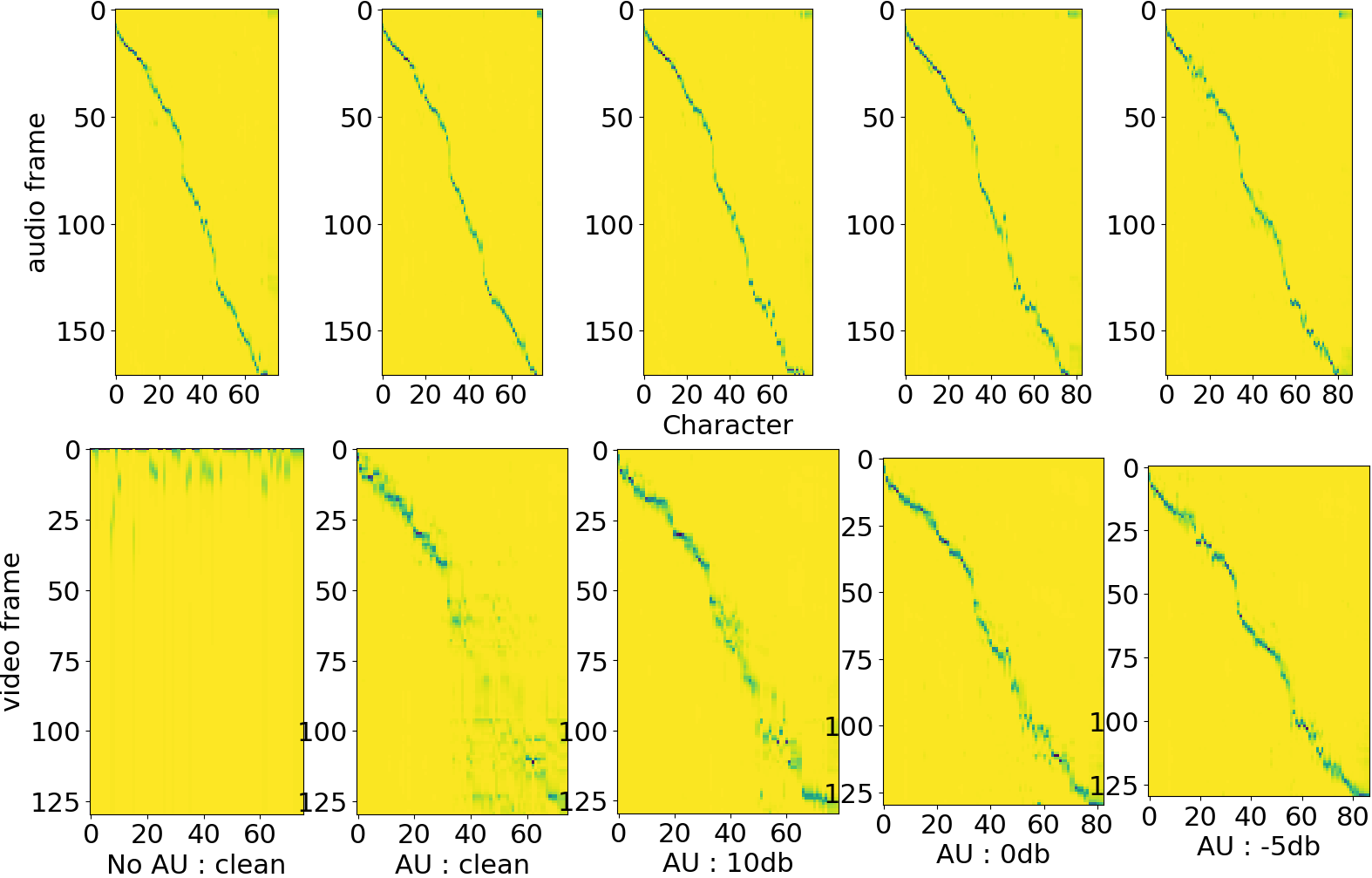}
        \caption{Alignment patterns of the WLAS network \cite{chung_cvpr_2017} trained on LRS2 and evaluated on the same sentence as in Figure~\ref{fig:exp4_fixed_alignments}. Top row displays the text to audio alignment, bottom row displays the text to video alignment.}
        \label{fig:exp10}
    \end{figure}

    We further investigate the two alignments produced by this architecture, consisting of the correlation of the decoder state with each encoded modality, which are displayed in Figure~\ref{fig:exp10}. The first column represents the \emph{AV Cat} system, presenting a similar video convergence problem as with the cross-modal alignment of \emph{AV Align} from Figure~\ref{fig:alignment_lrs2_av_faulty}. The next four columns illustrate the benefit of the AU loss for the video convergence of \emph{AV Cat + AU} as the noise level increases. The text to video alignments are less pronounced in clean speech conditions, unlike in Section~\ref{sec:lrs2_alignments}, where audio-visual alignments emerge and remain crisp starting on clean speech. This suggests that aligning the voice with the lips may be a simpler task than correlating characters with lips. In fact, the latter may prove difficult even to human annotators, making \emph{AV Align} more suitable for semi-supervised learning than \emph{AV Cat}.

\subsection{Error Analysis}

    In Section~\ref{exp:lrs2_results} we showed that \emph{AV Align + AU} performs 31\% better than the audio-only system on the test set of LRS2. Since this is only an average value, it would be interesting to know if this gain is general or restricted to a subset of the sentences. Therefore we have analysed the error rate on individual sentences. For each test sentence, we first compute the difference $\delta$ between the error rates of the \emph{Audio} and \emph{AV Align + AU} systems. Next, we estimate the predictability of each sentence by training a separate character-level language model on the train set, and evaluating the cross-entropy between the labels and the predictions. The language model is similar to the \emph{AV Align + AU} decoder shown in Figure~\ref{fig:av_align}, but without any conditioning on the encoder. In Table~\ref{tab:exp8} we list several examples of sentences ranked by their cross-entropy score.
    Since the analysis is fairly similar for all noise levels, we focus our attention on the most challenging -5db condition.
    
\begin{figure}[t]
        \centering
        \includegraphics[width=0.7\linewidth]{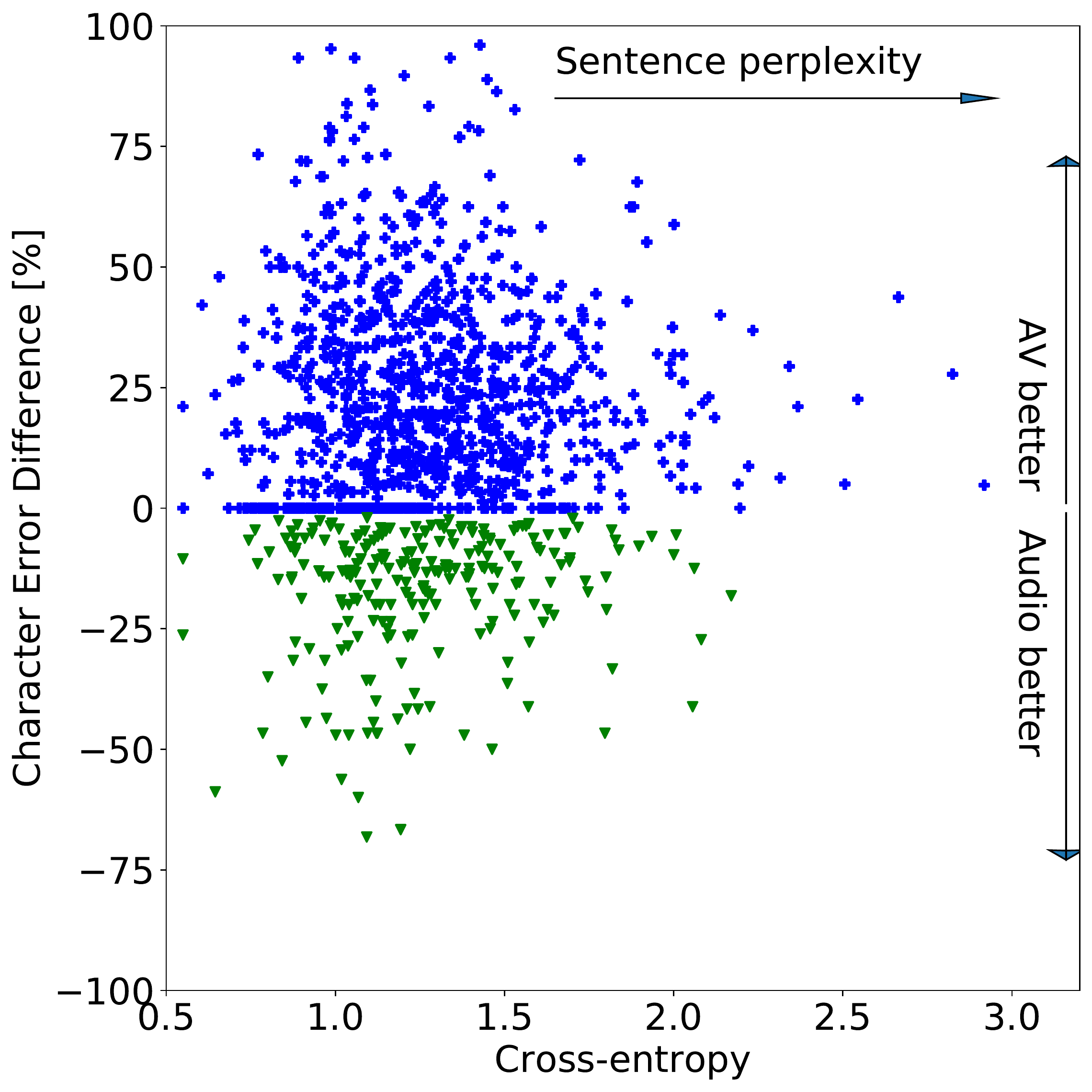}
        \caption{Absolute error difference between the Audio and \emph{AV Align + AU} systems on \mbox{-5db} speech, sorted by their predictability (easier sentences on left).}
        \label {fig:exp8}
    \end{figure}

\begin{figure}[t]
        \centering
        \includegraphics[width=\linewidth]{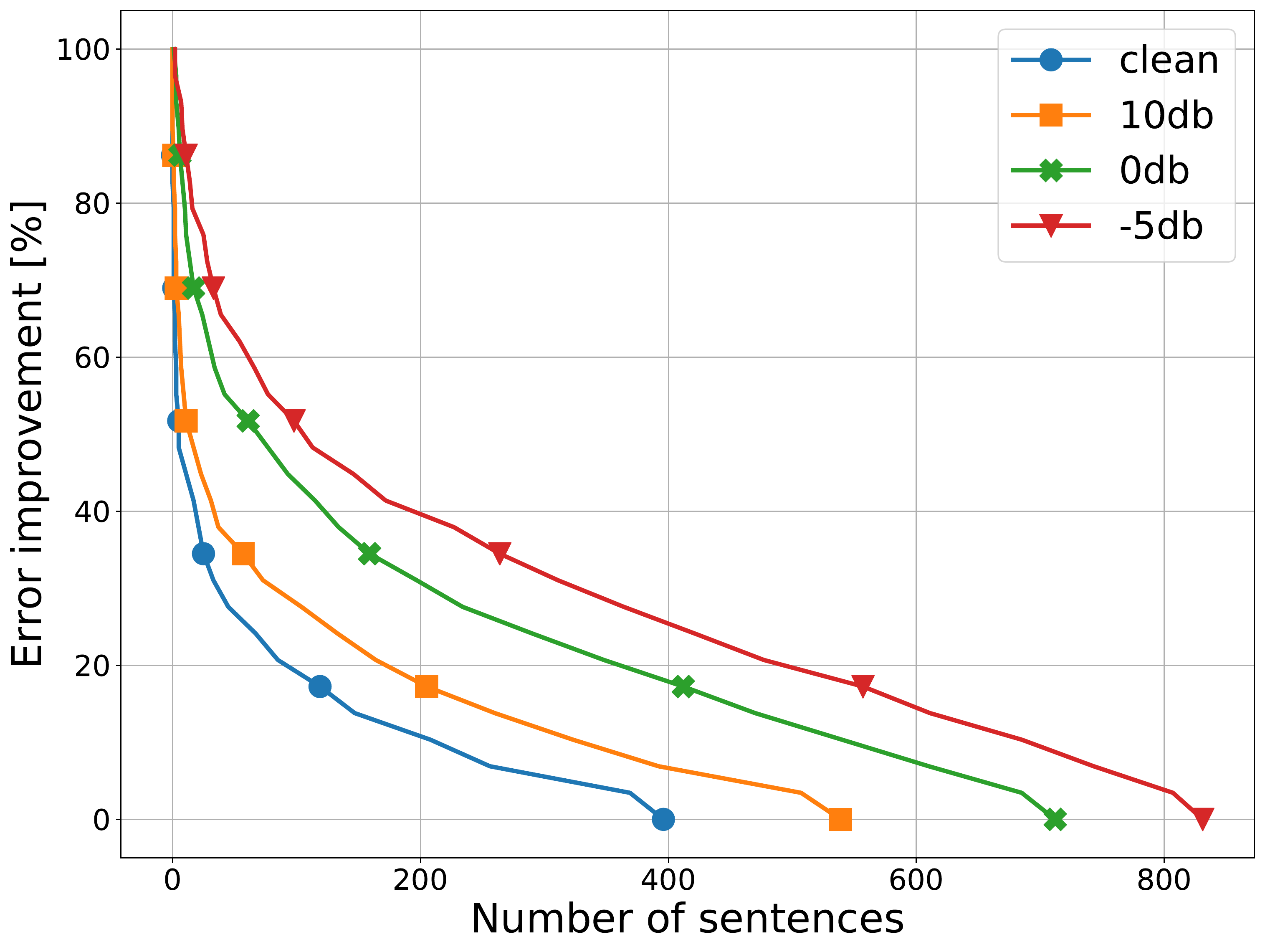}
        \caption{Cumulative distribution function of the error improvement on LRS2, truncated to 0\%, from \emph{Audio} to \emph{AV Align + AU}. The test set of LRS2 contained a number of \textbf{1,222} sentences. \emph{X sentences see an improvement of at least Y$\%$.}}
        \label{fig:exp8_cdf}
    \end{figure}

    We plot the error difference $\delta$ in Figure~\ref{fig:exp8}. Although \emph{AV Align + AU} performs better on average, there is still a number of sentences where the audio system scores better.
    A closer inspection on several examples where the error difference is -50\% or lower shows an interesting pattern: while the audio system makes reasonable spelling mistakes at this noise level, the prediction of the audio-visual one looks highly uncorrelated with the input. For example, the sentence "was it your choice" is acoustically transcribed as "was in your choice", whereas the \emph{AV Align + AU} prediction is "was in the auctions". This sentence belongs to a cluster of highly predictable sentences which are decoded almost perfectly by the audio system. We could not identify an obvious pattern in the visual domain on these sentences. We performed an analogous analysis between the audio and \emph{AV Cat + AU} systems, and also between \emph{AV Cat + AU} and \emph{AV Align + AU}, all with similar findings. This result suggests a shortcoming of both audio-visual systems: they do not fall back to audio-only performance when not able to capitalise on the visual modality. Instead, the conditioning on the input seems to diminish, leading to a more prominent impact of the intrinsic language model.
    In Figure~\ref{fig:exp8_cdf} we illustrate the same difference $\delta$ as a cumulative distribution function, allowing us to compare rate of improvement under different noise conditions. In line with the results in Figure~\ref{fig:lrs2_3}, we notice that more sentences see the benefit of the visual modality in worsening audio conditions.
    
    \begin{table}[t]
\setlength{\dashlinegap}{4.5pt}
\setlength{\dashlinedash}{0.5pt}
\renewcommand{\arraystretch}{1.5}
\centering
\scriptsize
\caption{Examples of Sentences from LRS2 Ranked by their Cross-Entropy (CE) Score Reflecting Predictability}
\label{tab:exp8}
\begin{tabular}{p{2cm}lp{2cm}p{2cm}}
Sentence & CE & [A] prediction & [AV] prediction \\ \hline
         squirrel pox virus                  &  3.11 & spiral pops fires     & squeer apots fires                   \\ \hdashline
         puerto rican style &      2.8            &    porture recan style              &     porture reconsile                    \\ \hdashline
         great leonard cohen & 2.43& rate leader cowin & rate lenent cowen \\ \hdashline
         sausages in bacon & 2.34 & such a years in baken & such a year's impainent \\ \hdashline
         the duke of gloucester & 2.00 & which you could prossed & the two coffloster \\ \hline
         there aren't any biscuits in that barrel & 1.78 & there are antique biscuits in their barrow &  there aren't any buscuity in their bear of\\ \hdashline
         some decent scores & 1.66 & some piece of scores & some things that's all \\ \hdashline
         was it your choice & 1.36 & was in your choice & was in the auctions \\ \hdashline
         very close by the university & 1.08 & very close by the university & very close by the university \\ \hdashline
         and our experts & 1.05 & i know where it's that & and our experts \\
         \hline
         i don't think so & 0.89 & i don't think so & but don't place so \\ \hdashline
         something like that &  0.63                &   something like that               &    something like that                     \\ \hdashline
         thank you very much &     0.56             &   thank you very much               &      thank you very much                  
\end{tabular}
\end{table}
    
\section{Discussion and Conclusions}
\label{sec:discuss}

\emph{AV Align} is the first architecture for AVSR that explicitly and automatically learns the asynchronous alignments between the audio and visual modalities of speech. We have demonstrated our results on two large publicly available datasets in the AVSR domain, and the code is publicly shared. This is an important result because it allows the system to capitalise on the visual modality without requiring pre-training strategies, while creating the opportunity to carry out phonetic investigations thanks to its interpretability property.
The system learns to discover audio-visual alignment patterns that provide informative visual cues to the audio modality, despite not being explicitly instructed to do so. This result is comparable with previous findings on traditional sequence to sequence neural networks learning the monotonicity of acoustic speech  \cite{chan_icassp2016, chorowski_neurips2015} or visual speech \cite{sterpu_icip2018, chung_cvpr_2017}, as the decoded graphemes align with their corresponding modality representations. However, before this work it had never been demonstrated that this property holds for the cross-modal alignment between two encoders.

Many researchers have encountered difficulties in capitalising on the visual modality of speech given a dominant acoustic one under low noise conditions. Common solutions resorted to pre-training the visual front-end on a different vision task, or to an alternation between the two modalities in training, where one of them is randomly disconnected when learning the rest of the parameters in the system. The interpretability properties of \emph{AV Align} have given us a greater insight into the nature of the optimisation problem, and motivated us to propose the regression of two lip-related Action Units from visual representations as a secondary objective. Our approach greatly simplifies the training strategy, enabling our system to achieve competitive error rate reductions with a fraction of the training data required by other approaches.

Finally, we make a direct comparison with the more popular audio-visual fusion scheme presented in \cite{chung_cvpr_2017}, although without making use of the full training procedure of WLAS. We show that such an approach can also benefit from the addition of the secondary AU loss, yet to a lesser extent than \emph{AV Align}, confirming the difficulty of learning good visual representations in AVSR. A closer look at the alignment patterns suggests that learning cross-modal correlations as in \emph{AV Align} may be a more suitable approach for AVSR than relating the state of the WLAS bimodal decoder to each modality separately.

A main take away message from error analysis is that the performance improvements reported with multimodal systems are affected by a high deviation from the mean, leading to a considerable number of sentences where the audio system is ahead by a large error margin. This exposes a fundamental challenge in AVSR, that the visual modality needs to be integrated without impairing the auditory one, which in turn may require mechanisms for assessing the confidence in the visual content. This may warrant a re-evaluation of approaches originally designed for HMM frameworks such as those of Papandreou~\etal~\cite{papandreou2009}.
Filtering out unreliable video sources may prove particularly important for challenging datasets such as LRS2, as our investigation suggests that neural networks have difficulties in learning this skill automatically. Future AVSR systems may need to be designed and tested with these observations in mind, so they could fall back to audio-only performance whenever the visual modality is not informative.

\begin{figure}[t]
    \centering
    \includegraphics[width=1.0\linewidth]{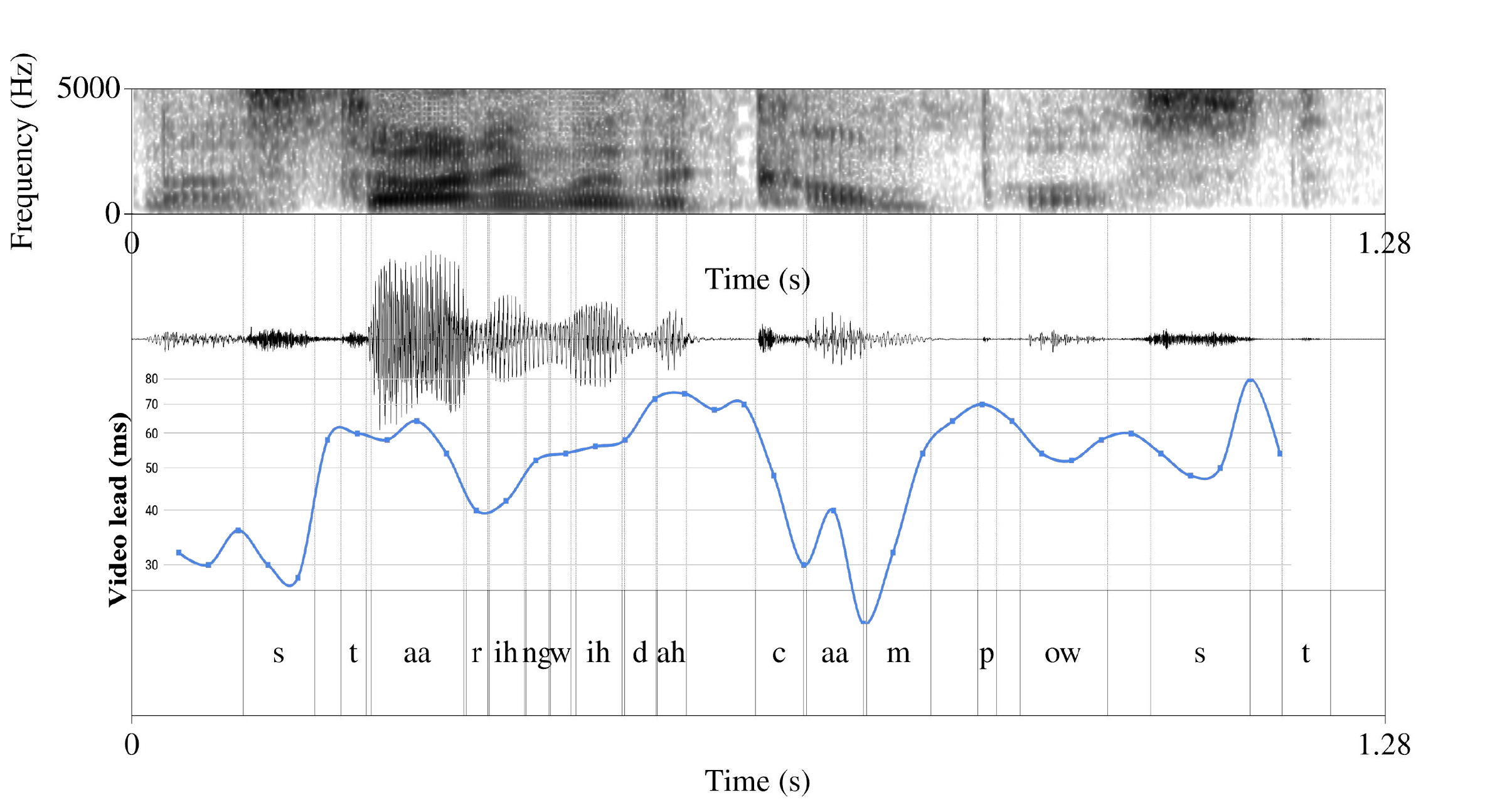}
    \caption{Phonetic analysis of the modality lags predicted by AV Align for the sentence \emph{"Starting with the compost"}, showing the speech spectrogram, waveform, modality lag, and transcription. The delay between modalities is estimated by fitting a normal distribution for each column (audio frame) of the cross-modal alignment matrix and selecting the mean.}
    \label{fig:phonetic_analysis}
\end{figure}

Overall, this work brings more evidence to support the idea of cross-modal alignment in AVSR, which has been largely overlooked so far. Despite having the entire sentence available for alignment, \emph{AV Align} learns to extract a visual context from a relatively narrow time window.
The estimated timing of this context vector, shown in Figure~\ref{fig:phonetic_analysis}, suggests that the learnt asynchronies between modalities vary between 20ms audio lead to 80ms video lead, with notable peaks associated with plosive sounds (\emph{t, d, p, t}). This is in line with the precise phonetic measurements of Schwartz and Savariaux~\cite{schwartz2014}, although a deeper analysis is needed to understand the learnt alignments, and we leave it as future work.


\section*{Acknowledgment}

We would like to thank Prof. Fran\c cois Piti\'e for his valuable advice on the statistical significance tests. Our work is supported by a GPU grant from NVIDIA. The ADAPT Centre for Digital Content Technology is funded under the SFI Research Centres Programme (Grant 13/RC/2106) and is co-funded under the European Regional Development Fund.

\bibliographystyle{IEEEtran}
\bibliography{mybib}

\begin{thebibliography}{10}
\providecommand{\url}[1]{#1}
\csname url@samestyle\endcsname
\providecommand{\newblock}{\relax}
\providecommand{\bibinfo}[2]{#2}
\providecommand{\BIBentrySTDinterwordspacing}{\spaceskip=0pt\relax}
\providecommand{\BIBentryALTinterwordstretchfactor}{4}
\providecommand{\BIBentryALTinterwordspacing}{\spaceskip=\fontdimen2\font plus
\BIBentryALTinterwordstretchfactor\fontdimen3\font minus
  \fontdimen4\font\relax}
\providecommand{\BIBforeignlanguage}[2]{{%
\expandafter\ifx\csname l@#1\endcsname\relax
\typeout{** WARNING: IEEEtran.bst: No hyphenation pattern has been}%
\typeout{** loaded for the language `#1'. Using the pattern for}%
\typeout{** the default language instead.}%
\else
\language=\csname l@#1\endcsname
\fi
#2}}
\providecommand{\BIBdecl}{\relax}
\BIBdecl

\bibitem{massaro1998}
D.~W. Massaro and D.~G. Stork, ``Speech recognition and sensory integration: A
  240-year-old theorem helps explain how people and machines can integrate
  auditory and visual information to understand speech,'' \emph{American
  Scientist}, vol.~86, no.~3, pp. 236--244, 1998.

\bibitem{fernandez2017}
A.~{Fernandez-Lopez}, O.~{Martinez}, and F.~M. {Sukno}, ``Towards estimating
  the upper bound of visual-speech recognition: The visual lip-reading
  feasibility database,'' in \emph{12th IEEE International Conference on
  Automatic Face Gesture Recognition}, May 2017, pp. 208--215.

\bibitem{oviatt2017}
S.~Oviatt, B.~Schuller, P.~R. Cohen, D.~Sonntag, G.~Potamianos, and
  A.~Kr\"{u}ger, Eds., \emph{The Handbook of Multimodal-Multisensor Interfaces:
  Foundations, User Modeling, and Common Modality Combinations}.\hskip 1em plus
  0.5em minus 0.4em\relax New York, NY, USA: ACM and Morgan \& Claypool, 2017,
  vol.~1.

\bibitem{lalor2018}
E.~C. Lalor, ``Neuroscience: The rhythms of speech understanding,''
  \emph{Current Biology}, vol.~28, no.~3, pp. R105 -- R108, 2018.

\bibitem{irwin2017}
J.~Irwin and L.~DiBlasi, ``Audiovisual speech perception: A new approach and
  implications for clinical populations,'' \emph{Language and Linguistics
  Compass}, vol.~11, no.~3, p. e12237, 2017, e12237 LNCO-0650.R2.

\bibitem{potamianos2003}
G.~Potamianos, C.~Neti, G.~Gravier, A.~Garg, and A.~W. Senior, ``Recent
  advances in the automatic recognition of audiovisual speech,''
  \emph{Proceedings of the IEEE}, vol.~91, no.~9, pp. 1306--1326, Sept 2003.

\bibitem{FERNANDEZLOPEZ2018}
A.~Fernandez-Lopez and F.~M. Sukno, ``Survey on automatic lip-reading in the
  era of deep learning,'' \emph{Image and Vision Computing}, vol.~78, pp. 53 --
  72, 2018.

\bibitem{Potamianos:2017}
G.~Potamianos, E.~Marcheret, Y.~Mroueh, V.~Goel, A.~Koumbaroulis,
  A.~Vartholomaios, and S.~Thermos, ``Audio and visual modality combination in
  speech processing applications,'' in \emph{{The Handbook of
  Multimodal-Multisensor Interfaces}}.\hskip 1em plus 0.5em minus 0.4em\relax
  New York, NY, USA: ACM and Morgan \& Claypool, 2017, pp. 489--543.

\bibitem{Lucey2005}
P.~Lucey, D.~Dean, and S.~Sridharan, ``Problems associated with current
  area-based visual speech feature extraction techniques,'' in
  \emph{Auditory-Visual Speech Processing (AVSP)}, 2005, pp. 73--78.

\bibitem{cox2008}
S.~Cox, R.~Harvey, Y.~Lan, J.~Newman, and B.~Theobald, ``The challenge of
  multispeaker lip-reading,'' in \emph{International Conference on
  Auditory-Visual Speech Processing}, September 2008, pp. 179--184.

\bibitem{attention_seq2seq}
D.~Bahdanau, K.~Cho, and Y.~Bengio, ``Neural machine translation by jointly
  learning to align and translate,'' in \emph{3rd International Conference on
  Learning Representations, {ICLR} 2015, San Diego, CA, USA, May 7-9, 2015,
  Conference Track Proceedings}, 2015, pp. 1--15.

\bibitem{transformers}
A.~Vaswani, N.~Shazeer, N.~Parmar, J.~Uszkoreit, L.~Jones, A.~N. Gomez,
  L.~Kaiser, and I.~Polosukhin, ``Attention is all you need,'' in
  \emph{Advances in Neural Information Processing Systems 30}, 2017, pp.
  5998--6008.

\bibitem{lstm}
F.~Gers, J.~Schmidhuber, and F.~Cummins,
  ``\BIBforeignlanguage{English}{Learning to forget: continual prediction with
  {LSTM}},'' \emph{\BIBforeignlanguage{English}{IET Conference Proceedings}},
  pp. 850--855, January 1999.

\bibitem{petridis_ctc_2018}
S.~{Petridis}, T.~{Stafylakis}, P.~{Ma}, G.~{Tzimiropoulos}, and M.~{Pantic},
  ``Audio-visual speech recognition with a hybrid ctc/attention architecture,''
  in \emph{2018 IEEE Spoken Language Technology Workshop (SLT)}, Dec 2018, pp.
  513--520.

\bibitem{afouras_pami}
T.~{Afouras}, J.~S. {Chung}, A.~{Senior}, O.~{Vinyals}, and A.~{Zisserman},
  ``Deep audio-visual speech recognition,'' \emph{IEEE Transactions on Pattern
  Analysis and Machine Intelligence}, pp. 1--1, 2018.

\bibitem{ICML2011Ngiam_399}
J.~Ngiam, A.~Khosla, M.~Kim, J.~Nam, H.~Lee, and A.~Ng, ``Multimodal deep
  learning,'' in \emph{Proceedings of the 28th International Conference on
  Machine Learning (ICML-11)}.\hskip 1em plus 0.5em minus 0.4em\relax ACM, June
  2011, pp. 689--696.

\bibitem{chung_cvpr_2017}
J.~Son~Chung, A.~Senior, O.~Vinyals, and A.~Zisserman, ``Lip reading sentences
  in the wild,'' in \emph{The IEEE Conference on Computer Vision and Pattern
  Recognition (CVPR)}, July 2017, pp. 3444--3453.

\bibitem{zhou2018}
P.~{Zhou}, W.~{Yang}, W.~{Chen}, Y.~{Wang}, and J.~{Jia}, ``Modality attention
  for end-to-end audio-visual speech recognition,'' in \emph{ICASSP 2019 - 2019
  IEEE International Conference on Acoustics, Speech and Signal Processing
  (ICASSP)}, May 2019, pp. 6565--6569.

\bibitem{shillingford2018}
B.~Shillingford, Y.~M. Assael, M.~W. Hoffman, T.~Paine, C.~Hughes, U.~Prabhu,
  H.~Liao, H.~Sak, K.~Rao, L.~Bennett, M.~Mulville, B.~Coppin, B.~Laurie, A.~W.
  Senior, and N.~de~Freitas, ``Large-scale visual speech recognition,''
  \emph{CoRR}, vol. abs/1807.05162, 2018.

\bibitem{Sterpu_ICMI2018}
G.~Sterpu, C.~Saam, and N.~Harte, ``{Attention-based Audio-Visual Fusion for
  Robust Automatic Speech Recognition},'' in \emph{Proceedings of the 20th ACM
  International Conference on Multimodal Interaction}, ser. ICMI '18.\hskip 1em
  plus 0.5em minus 0.4em\relax New York, NY, USA: ACM, 2018, pp. 111--115.

\bibitem{ekman1997face}
P.~Ekman and E.~L. Rosenberg, \emph{What the face reveals: Basic and applied
  studies of spontaneous expression using the Facial Action Coding System
  (FACS)}.\hskip 1em plus 0.5em minus 0.4em\relax Oxford University Press, USA,
  1997.

\bibitem{lrs2}
BBC and O.~University, ``{The Oxford-BBC Lip Reading Sentences 2 (LRS2)
  Dataset},'' \url{http://www.robots.ox.ac.uk/~vgg/data/lip_reading/lrs2.html},
  2017, online, Accessed: 7 June 2019.

\bibitem{tcdtimit}
N.~Harte and E.~Gillen, ``{TCD-TIMIT}: An audio-visual corpus of continuous
  speech,'' \emph{IEEE Transactions on Multimedia}, vol.~17, no.~5, pp.
  603--615, May 2015.

\bibitem{arpit2017}
D.~Arpit, S.~Jastrzkebski, N.~Ballas, D.~Krueger, E.~Bengio, M.~S. Kanwal,
  T.~Maharaj, A.~Fischer, A.~Courville, Y.~Bengio, and S.~Lacoste-Julien, ``A
  closer look at memorization in deep networks,'' in \emph{Proceedings of the
  34th International Conference on Machine Learning - Volume 70}, ser.
  ICML'17.\hskip 1em plus 0.5em minus 0.4em\relax JMLR.org, 2017, pp. 233--242.

\bibitem{zhang2017}
C.~Zhang, S.~Bengio, M.~Hardt, B.~Recht, and O.~Vinyals, ``Understanding deep
  learning requires rethinking generalization,'' in \emph{International
  Conference on Learning Representations}, 2017, pp. 1--15.

\bibitem{petridis_icassp2018}
S.~{Petridis}, T.~{Stafylakis}, P.~{Ma}, F.~{Cai}, G.~{Tzimiropoulos}, and
  M.~{Pantic}, ``End-to-end audiovisual speech recognition,'' in \emph{2018
  IEEE International Conference on Acoustics, Speech and Signal Processing
  (ICASSP)}, April 2018, pp. 6548--6552.

\bibitem{baltrusaitis2019}
T.~{Baltrušaitis}, C.~{Ahuja}, and L.~{Morency}, ``Multimodal machine
  learning: A survey and taxonomy,'' \emph{IEEE Transactions on Pattern
  Analysis and Machine Intelligence}, vol.~41, no.~2, pp. 423--443, Feb 2019.

\bibitem{purwins2019}
H.~{Purwins}, B.~{Li}, T.~{Virtanen}, J.~{Schluter}, S.~{Chang}, and T.~N.
  {Sainath}, ``Deep learning for audio signal processing,'' \emph{IEEE Journal
  of Selected Topics in Signal Processing}, pp. 1--1, 2019.

\bibitem{schwartz2014}
J.-L. Schwartz and C.~Savariaux, ``No, there is no 150 ms lead of visual speech
  on auditory speech, but a range of audiovisual asynchronies varying from
  small audio lead to large audio lag,'' \emph{PLOS Computational Biology},
  vol.~10, no.~7, pp. 1--10, 07 2014.

\bibitem{karpov2011}
A.~Karpov, A.~Ronzhin, I.~S. Kipyatkova, and M.~Zelezny, ``Influence of
  phone-viseme temporal correlations on audiovisual {STT} and {TTS}
  performance,'' in \emph{17th International Congress of Phonetic Sciences,
  ICPhS 2011, Hong Kong, China, August 17-21}, 2011, pp. 1030--1033.

\bibitem{tao_icme2018}
F.~{Tao} and C.~{Busso}, ``Aligning audiovisual features for audiovisual speech
  recognition,'' in \emph{2018 IEEE International Conference on Multimedia and
  Expo (ICME)}, July 2018, pp. 1--6.

\bibitem{Stafylakis2017}
T.~Stafylakis and G.~Tzimiropoulos, ``Combining residual networks with lstms
  for lipreading,'' in \emph{Interspeech}, 2017, pp. 3652--3656.

\bibitem{tao_taslp2018}
F.~{Tao} and C.~{Busso}, ``Gating neural network for large vocabulary
  audiovisual speech recognition,'' \emph{IEEE/ACM Transactions on Audio,
  Speech, and Language Processing}, vol.~26, no.~7, pp. 1290--1302, 2018.

\bibitem{He2019}
K.~He, R.~Girshick, and P.~Dollar, ``Rethinking imagenet pre-training,'' in
  \emph{The IEEE International Conference on Computer Vision (ICCV)}, October
  2019, pp. 4917--4926.

\bibitem{resnet2}
K.~He, X.~Zhang, S.~Ren, and J.~Sun, ``Identity mappings in deep residual
  networks,'' in \emph{ECCV 2016}.\hskip 1em plus 0.5em minus 0.4em\relax
  Springer International, 2016, pp. 630--645.

\bibitem{Thangthai2017}
K.~Thangthai and R.~Harvey, ``Improving computer lipreading via dnn sequence
  discriminative training techniques,'' in \emph{Interspeech}, 2017, pp.
  3657--3661.

\bibitem{sterpu2017}
G.~Sterpu and N.~Harte, ``Towards lipreading sentences using active appearance
  models,'' in \emph{AVSP}, Stockholm, Sweden, August 2017, pp. 70--75.

\bibitem{Thangthai2018}
K.~Thangthai and R.~Harvey, ``Building large-vocabulary speaker-independent
  lipreading systems,'' in \emph{Interspeech}, 2018, pp. 2648--2652.

\bibitem{assael2016}
Y.~M. Assael, B.~Shillingford, S.~Whiteson, and N.~de~Freitas, ``Lipnet:
  Sentence-level lipreading,'' \emph{CoRR}, vol. abs/1611.01599, 2016.

\bibitem{sterpu_icip2018}
G.~Sterpu, C.~Saam, and N.~Harte, ``{Can DNNs Learn to Lipread Full
  Sentences?}'' in \emph{2018 25th IEEE International Conference on Image
  Processing (ICIP)}, Oct 2018, pp. 16--20.

\bibitem{chung2016}
J.~S. Chung and A.~Zisserman, ``Lip reading in the wild,'' in \emph{Computer
  Vision -- ACCV 2016}, S.-H. Lai, V.~Lepetit, K.~Nishino, and Y.~Sato,
  Eds.\hskip 1em plus 0.5em minus 0.4em\relax Cham: Springer International
  Publishing, 2017, pp. 87--103.

\bibitem{wand2016}
M.~{Wand}, J.~{Koutník}, and J.~{Schmidhuber}, ``Lipreading with long
  short-term memory,'' in \emph{2016 IEEE International Conference on
  Acoustics, Speech and Signal Processing (ICASSP)}, March 2016, pp.
  6115--6119.

\bibitem{Stafylakis2018}
T.~{Stafylakis} and G.~{Tzimiropoulos}, ``Deep word embeddings for visual
  speech recognition,'' in \emph{IEEE International Conference on Acoustics,
  Speech and Signal Processing (ICASSP)}, April 2018, pp. 4974--4978.

\bibitem{bear2017}
H.~L. Bear and S.~L. Taylor, ``Visual speech recognition: aligning
  terminologies for better understanding,'' in \emph{Proceedings of British
  Machine Vision Conference}.\hskip 1em plus 0.5em minus 0.4em\relax BMVA
  Press, September 2017, pp. 1--11.

\bibitem{hori2017}
C.~{Hori}, T.~{Hori}, T.~{Lee}, Z.~{Zhang}, B.~{Harsham}, J.~R. {Hershey},
  T.~K. {Marks}, and K.~{Sumi}, ``Attention-based multimodal fusion for video
  description,'' in \emph{2017 IEEE International Conference on Computer Vision
  (ICCV)}, Oct 2017, pp. 4203--4212.

\bibitem{abadi2016tensorflow}
M.~Abadi, P.~Barham, J.~Chen, Z.~Chen, A.~Davis, J.~Dean, M.~Devin,
  S.~Ghemawat, G.~Irving, M.~Isard \emph{et~al.}, ``Tensorflow: a system for
  large-scale machine learning.'' in \emph{OSDI}, vol.~16, 2016, pp. 265--283.

\bibitem{luong_attention}
T.~Luong, H.~Pham, and C.~D. Manning, ``Effective approaches to attention-based
  neural machine translation,'' in \emph{Proceedings of the 2015 Conference on
  Empirical Methods in Natural Language Processing}.\hskip 1em plus 0.5em minus
  0.4em\relax Association for Computational Linguistics, 2015, pp. 1412--1421.

\bibitem{openface2}
T.~Baltrusaitis, A.~Zadeh, Y.~C. Lim, and L.~Morency, ``Openface 2.0: Facial
  behavior analysis toolkit,'' in \emph{13th IEEE International Conference on
  Automatic Face Gesture Recognition}, May 2018, pp. 59--66.

\bibitem{jeffers1980speechreading}
J.~Jeffers and M.~Barley, \emph{Speechreading (lipreading)}.\hskip 1em plus
  0.5em minus 0.4em\relax Charles C. Thomas Publisher, 1980.

\bibitem{garofalo1993darpa}
J.~S. Garofalo, L.~F. Lamel, W.~M. Fisher, J.~G. Fiscus, D.~S. Pallett, and
  N.~L. Dahlgren, ``The darpa timit acoustic-phonetic continuous speech corpus
  cdrom,'' \emph{Linguistic Data Consortium}, vol. LDC93S1, pp. 1--94, 1993.

\bibitem{chan_icassp2016}
W.~Chan, N.~Jaitly, Q.~Le, and O.~Vinyals, ``Listen, attend and spell: A neural
  network for large vocabulary conversational speech recognition,'' in
  \emph{2016 IEEE International Conference on Acoustics, Speech and Signal
  Processing (ICASSP)}, March 2016, pp. 4960--4964.

\bibitem{chorowski_neurips2015}
J.~Chorowski, D.~Bahdanau, D.~Serdyuk, K.~Cho, and Y.~Bengio, ``Attention-based
  models for speech recognition,'' in \emph{Proceedings of the 28th
  International Conference on Neural Information Processing Systems}.\hskip 1em
  plus 0.5em minus 0.4em\relax MIT Press, 2015, pp. 577--585.

\bibitem{papandreou2009}
G.~{Papandreou}, A.~{Katsamanis}, V.~{Pitsikalis}, and P.~{Maragos}, ``Adaptive
  multimodal fusion by uncertainty compensation with application to audiovisual
  speech recognition,'' \emph{IEEE Transactions on Audio, Speech, and Language
  Processing}, vol.~17, no.~3, pp. 423--435, March 2009.

\end{thebibliography}

\end{document}